\documentstyle[twocolumn,aps,prd,epsf]{revtex}
\begin{document}
\draft
\title{A New Perspective on the Scalar Meson Puzzle, from
Spontaneous Chiral Symmetry Breaking Beyond BCS}
\author{Pedro J. de A. Bicudo}
\address{
Departamento de F\'\i sica and
Centro de F\'\i sica das Interac\c c\~oes Fundamentais,
Edif\'\i cio Ci\^encia, Instituto Superior T\'ecnico,
Av. Rovisco Pais, 1096 Lisboa, Portugal
}
%\twocolumn[ 
\maketitle
%\widetext 
\begin{abstract} 
We introduce coupled channels of Bethe-Salpeter
mesons both in the boundstate equation for mesons
and in the mass gap equation for chiral symmetry.
Consistency is insured by the Ward Identities for axial
currents, which preserve the Goldstone boson nature of the pion
and prevents a systematic shift of the hadron spectrum.
We study the decay of a scalar meson coupled to a pair
of pseudoscalars.
We also show that coupled channels reduce the breaking of chiral
symmetry, with the same Feynman diagrams that appear in the
coupling of a scalar meson to a pair of pseudoscalar mesons.
Exact calculations are performed in a particular confining
quark model, where we find that the groundstate $I=0, \ {}^3P_0$
$q \bar q$ meson is the $f_0(980)$ with a partial decay
width of $40MeV$.
We also find a $30 \%$ reduction of the chiral condensate
due to coupled channels.
\end{abstract}
%]                     %
%\narrowtext           %
%
%sssssssssssssssssssssssssssssssssssssssssssssssssssssssssssssssssssssssss
\section{Introduction}
%
%uuuuuuuuuuuuuuuuuuuuuuuuuuuuuuuuuuuuuuuuuuuuuuuuuuuuuuuuuuuuuuuuuuuuuuuuu
\subsubsection*{The Scalar puzzle}
The scalar mesons form perhaps the most puzzling family in hadronic
physics.
The first puzzling fact concerns the experimental errors in the
partial decay widths, the decay widths and even the masses. The
lightest scalar, $f_0(400-1200)$, has a poorly determined mass.
The confidence on the decay widths of the $f_0(980)$ and $a_0(980)$
has also decreased \cite{PDB} strongly since 1994.
\par
\begin{figure}
\begin{picture}(248,115)(0,0)
\put(0,0){\epsffile{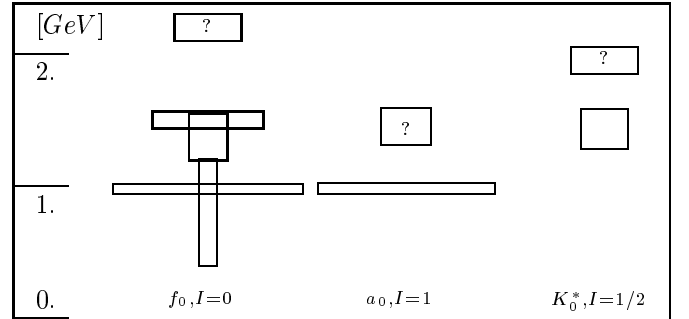}}\end{picture}
\caption{ We represent,
with rectangles of height $\Gamma$,
the experimentally observed scalar resonances
according to the Review of Particle Properties,
and mark the unconfirmed ones with a ''?''.
In a naive quark model,
the $a_0$ and $K^*_0$ channels
should show half of the resonances of the $f_0$ channel.
\label{scalarPDB}}
\end{figure}

The other puzzling argument concerns the matching of the
the nine observed states with simple $SU(3)_f$ $q \bar q$ states.
One would expect four different towers of states corresponding
to the two $I=0 \ f_0$
(which are not degenerate for instance because the quark
$s$ has a clearly larger mass than the quarks $u, \, d$)
and to the $I=1 \ a_0$ , and the $I=1/2 \ K_0^*$.
A short glance at
Fig. \ref{scalarPDB} is sufficient to discard the single light
and extremely broad state, the $f_0(400-1200)$ as a simple member
of this family.
Then for the groundstates we could ascribe the narrowest states
which are respectively the $a_0(980)$,  $K_0^*(1430)$,
$f_0(980)$ and $f_0(1500)$, and for the radial excited states
we could respectively ascribe the $a_0(1450)$, $K_0^*(1950)$,
$f_0(1370)$ and $f_0(2200)$.
However the decay widths $\Gamma$ of the groundstate scalars
are narrower than expected when compared with other resonances
decaying in the same pseudoscalar pairs but with higher angular
momentum, except for the only precisely measured one, the
$K_0^*(1430)$.
Moreover the breaking of $SU(3)_f$ due to the $m_u\simeq m_d << m_s$
mass difference is nearly the double than expected when compared with
the vector meson family and with most baryons. This is only comparable
with the splittings in the pseudoscalar family.
%fffffffffffffffffffffffffffffffffffffffffffffffffffffffffff
%ffffffffffffffffffffffffffffffffffffffffffffffffffffffffff
\par
The scalar family is also the most interesting place to search for
the lightest (S-wave) non $q \bar q$ states. There are several
theoretical candidates to extra states which may be found.
The lightest glueball, which is expected from QCD, should be a
scalar \cite{Swanson}.
The model of Isgur and Weinstein et al \cite{Weinstein} suggests that
the narrowest scalars are meson-meson molecules.
The One Meson Exchange Potential models for the $NN$ interaction
usually postulates a scalar meson $\sigma$ with
a light $M\simeq 0.5 GeV$. 
In strongly coupled effective meson models \cite{Dullemond,Torsigma},
extra poles appear in the S matrix when couplings are large.
These meson models turn out to be the most successful models ones so 
far, explaining with complex nonlinear effects not only the narrow
$a_0(980)$ and $f_0(980)$ which are due to the vicinity of the
$K K$ threshold, but also the very wide and light $f_0(400-1200)$.
%
%uuuuuuuuuuuuuuuuuuuuuuuuuuuuuuuuuuuuuuuuuuuuuuuuuuuuuuuuuuuuuuuuuuuuuuuuuu
\subsubsection*{The relevance of chiral symmetry breaking}
Chiral symmetry breaking is important for the study of scalar
mesons and their decays for several reasons.
Unlike the vector, axial and tensor mesons, the scalar and pseudoscalar
mesons are mixed by the chiral rotations,
\begin{eqnarray}
\bar \psi \psi \rightarrow \cos(\theta) \bar \psi \psi
+ i \sin(\theta) \bar \psi \gamma_5 \psi \nonumber \\
\bar \psi \gamma_5 \psi \rightarrow -i \sin(\theta) \bar \psi \psi
+ \cos(\theta) \bar \psi \gamma_5 \psi \, ,
\end{eqnarray}
thus scalars and pseudoscalars are particularly sensitive to the chiral
symmetry.
The very small mass of the pseudoscalars is usually explained with chiral
symmetry breaking.
Moreover, since scalars decay essentially in pseudoscalars,
the pseudoscalar mass is important for the scalar decay.
Thus we expect that not only the pseudoscalar mesons but also
the scalars must contain the signature of the breaking of
chiral symmetry.
Inversely, the breaking of chiral symmetry is generated from the
trivial vacuum by scalar condensation, and we also expect that
the scalar properties should affect the breaking of chiral symmetry.
\par
These effect have been studied in phenomenological meson sigma models,
see for instance \cite{Torsigma}.
At the more microscopic level of quarks,
Dynamical Spontaneous Chiral Symmetry Breaking ($\chi S B$)
has been worked out in the past with several different quark-quark
effective interactions, at the same level as Bardeen
Cooper and Schrieffer (BCS) did \cite{BCS} for Superconductivity.
Since \cite{Nambu} Nambu and Jona-Lasinio (NJL) and until recently
\cite{bBCS in CSB,preprint} the mass gap
equation in chiral physics has been so far of the BCS type,
\cite{Nambu} including only the first order contribution from the
quark-quark interaction.
In this case the quark condensate  consists
\cite{3P0} of scalar $^3P_0$ quark antiquark pairs.
At the BCS level, which is very consistent, it is possible to derive
the mass gap equation in several different but equivalent methods.
The pseudoscalar
meson properties have been studied in great detail and the full meson
spectrum has been also been calculated in the literature\cite{papfpi}.
However the BCS approach is not exact in the case where coupled
channels of mesons are included.
%
%uuuuuuuuuuuuuuuuuuuuuuuuuuuuuuuuuuuuuuuuuuuuuuuuuuuuuuuuuuuuuuuuuuuuuu
\subsubsection*{Coupled Channels of Mesonic $q \bar q$ pairs}
\begin{figure}
\begin{picture}(165,40)(0,0)
\put(0,0){\epsffile{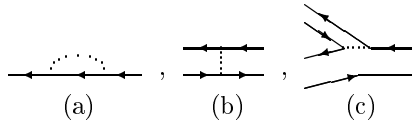}}\end{picture}
\caption{ We show the quartic diagrams which may contribute to
the boundstate equation.
In the strong coupling BCS, $(a)$ is included in the self energy
and $(b)$ is included in the interaction kernel.
Diagram $(c)$ which creates or annihilates quark-antiquark pairs
is only used in what we call beyond BCS.
\label{quartic diagrams}}
\end{figure}

In the case of weak coupled channel effects, it would be acceptable
to start from boundstates obtained at the BCS level and couple them
with the help of the annihilation diagrams of Fig.
\ref{quartic diagrams}, without changing the mass gap equation.
In this sense we started
some years ago to develop a program \cite{ourcouca}
to study the coupled channel effects in quark models
with chiral symmetry breaking.
%fffffffffffffffffffffffffffffffffffffffffffffffffffffffffff
%ffffffffffffffffffffffffffffffffffffffffffffffffffffffffffff
\par
The first result of our program was to reproduce \cite{ourcouca}
at the BCS level the strong decay of the vector meson $\rho$ (and
of the $\phi$). This has been studied by other authors recently
\cite{newrho}.
Later we extended our program to the nucleon interactions and had
good results \cite{ourcouca} in the $K_N$ s-wave scattering, the
$F_{N \pi N}$ and $F_{N \pi \Delta}$ derivative couplings and
the $N \, N$ short range interaction.
\par
However there is a recent trend in the literature to reevaluate coupled
channel effects in many hadronic phenomena. Some years ago they were
not supposed to account for more than $10\%$ of a hadron mass but
presently they are supposed to contribute with a negative mass shift
of the order of $50\%$ of the bare mass \cite{Richard,Thomas,Tornqvist}.
\par
Moreover it is possible to prove that the vacuum solution of the BCS
mass gap equation is not exact when coupled channels are included.
Suppose that the mass gap equation for chiral symmetry breaking was
solved at the BCS level, i.e. without including the coupled channels,
then a bare pion with vanishing bare mass would be found. If the coupled
channels were then included, at posteriori, in the bound state equation
then the pion mass would be the sum of the small bare mass plus a
mass shift and thus would have a resulting nonvanishing mass,
which implies that the pion lost its Goldstone boson nature.
This result is unavoidable, and can be proved variationally.
Thus the mass gap should be solved beyond BCS, especially
when scalar mesons are studied.
\par
\begin{figure}
\begin{picture}(227,95)(0,0)
\put(0,0){\epsffile{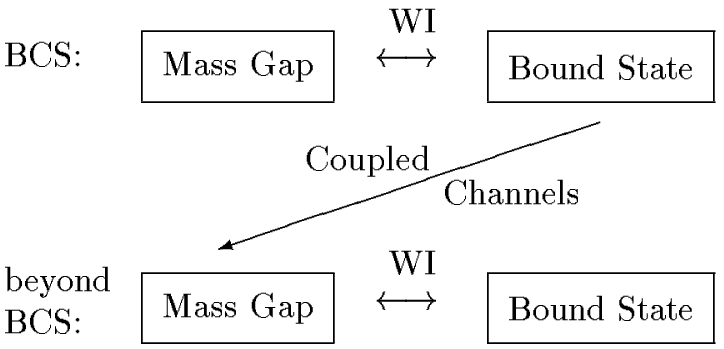}}\end{picture}
\caption{We illustrate the principle which is followed in this
paper to include the coupled channels in the mass gap equation
\label{prescription}}
\end{figure}

Solving this problem is a corner stone of our program.
It amounts to join the BCS mechanism with the mean field expansion 
of effective mesons.The logical path of the method we will follow is 
illustrated in Fig. \ref{prescription}. Self consistency is
insured by Ward Identities, and the scalar-pseudoscalar 
coupling will turn out to be crucial for this development.
The reward of solving chiral symmetry breaking with coupled channels
is a $\pi$ with a vanishing positive mass in the chiral limit
complying with all the theorems of PCAC, and a tower of resonances
(including the scalar meson resonances) above the $\pi$ with
higher masses due to radial, angular, or spin excitations.
In particular the resonances also have an imaginary component of
the mass $-i \Gamma \over 2$ that describes the decay width into
the open channels. 
%fffffffffffffffffffffffffffffffffffffffffffffffffffffffffff
%ffffffffffffffffffffffffffffffffffffffffffffffffffffffffff
%
%uuuuuuuuuuuuuuuuuuuuuuuuuuuuuuuuuuuuuuuuuuuuuuuuuuuuuuuuuuuu
\subsubsection*{The paper}
The aim of this paper is to study at the quark level some
meson decays which were studied in the literature without including
directly the full quark contributions \cite{Dullemond,Torsigma}.
We specialize in the groundstate $f_0 \rightarrow \pi \ \pi$ decay.
We also study the effect of the meson coupled channels on the
quark condensed vacuum.
\par
The remaining of this paper is organized as follows.
In section II we review the scalar masses and decays at the BCS level.
This includes the choice of an effective interaction for quarks,
the mass gap equation and the Bethe Salpeter equation at the BCS level,
the scalar coupling to pseudoscalars, and the scalar decays width.
In Section III we produce a finite extension beyond BCS for a class
of confining effective interactions, derive the mass gap equation
with coupled channels, link it to the scalar-pseudoscalars coupling,
and solve the mass gap equation.
Results are shown in Section IV together with their discussion.
We also include 4 appendices.
%ssssssssssssssssssssssssssssssssssssssssssssssssssssssssssssssssssssssssss
%
\section{The BCS level for a particular formalism}
%
%sssssssssssssssssssssssssssssssssssssssssssssssssssssssssssssssssssssssssss
\subsection{The choice of an effective interaction}
The quantitative results of this paper will be obtained with a
particular chiral invariant strong potential
which is an extended version of the NJL potential \cite{Nambu}.
For the study of dynamical $\chi S B$ it is crucial to have
a closed model where calculations can be carried until the end,
because precise cancellations occur.
At this point we abandon the explicit SU(3) gauge invariance.
We start by introducing a class \cite{papfpi,charm}
of Dirac quark Hamiltonians which are, in
the limit of massless quarks, explicitly chiral invariant,
\FL
\begin{eqnarray}
\label{hamilt}
H&=&\int\, d^3x \left[ \psi^{\dag}( x) \;(m_0\beta -i{\vec{\alpha}
\cdot \vec{\nabla}} )\;\psi( x)\;+\sum_{l}\right.
\nonumber \\
&&\left. { 1\over 2}
\overline{\psi}( x)
\Omega_l \psi ( x) \int \!d^4y  \; \:V_l( x -y)
\;\overline{\psi}( y)
\Omega_l
{2} \psi( y) \right]
\end{eqnarray}
The quark-quark interaction,  is an effective color dependent
2-body interaction.
In eq.~(\ref{hamilt}) the operators $\Omega_l$ include both
the color Gell-Mann matrices and the Dirac matrices.
The sum in the Dirac matrices must be chiral invariant.
Because the Gell-Mann matrices are traceless there will
be no tadpoles in this scheme.
In Hadronic Physics the effective interaction should be simultaneously 
color confining, in the Minkowsky space, local, and Lorentz invariant. 
However no interaction which complies with all these constraints has yet
been used to study chiral symmetry breaking.
\par
The models used in the literature divide essentially in two classes,
in particular the more popular one springs directly from NJL
\cite{Nambu}, is Euclidean and due to the structure of the interaction
has usually no analytic continuation to the Minkowsky space and lacks
confinement. This class has been extended in many different directions.
For instance finite size boundstates were included with the Global
Color Model of ref. \cite{Roberts}, and a sophisticated interaction
with a general tensor structure, an almost linear long range and
pertubative short range is found in ref. \cite{Liu}.
\par
The other class of models \cite{3P0,papfpi,Adler}
has the single drawback of using an instantaneous potential
(except for Lorentz invariant extensions \cite{Kalinowsky}
of this class).
But it has the advantage of being confining which
allows to study the whole 
\cite{papfpi,ourcouca,charm,Gastao,meson,prl} hadron spectrum.
This approximation also has the advantage to allow a straightforward
application to low energy nuclear physics.
For the theoretical foundations of these models, including the
connection to both pertubative and nonpertubative QCD, see
\cite{Lagae}.
\par
Thus we choose to calculate the quantitative results of this paper
within the second class of Nambu and Jona-Lasinio potentials.
We use a simple model which is in very
good agreement with the experiments in what concerns the hadronic
spectroscopy, the decays of the vector mesons $\rho$ and $\Phi$,
the coupling of a $\pi$ to a $N$ or $\Delta$ and the $N \ N$
short range interaction \cite{ourcouca}, moreover it supports that
chiral symmetry breaking is very stable in the presence of
Nuclear Matter \cite{prl}.
While confinement is an essential physical aspect of the model, the
instantaneous approximation simplifies drastically the energy dependence
of the interaction, and allows to work in a framework which is
familiar to Schr\"odinger' s equations.
The 2-body potential for Dirac quarks is,
\begin{equation}
\label{potential}
 {-3 \over 4} {\vec{\lambda} \over 2} \cdot \otimes
{\vec{\lambda} \over 2}
\left[ \gamma_0 \otimes \gamma_0 \left( k^3_0 r^2-U \right) +
a \vec \gamma \cdot \otimes \vec\gamma k^3_0 r^2 \right]\delta(t)
\end{equation}
and the Fourier transform of the potential is,
\begin{eqnarray}
\label{interaction}
\Omega_l\widetilde{V}_l(k)\otimes \Omega_l&=&
{-3 \over 4} {\vec{\lambda} \over 2} \cdot \otimes {\vec{\lambda} \over 2}
\left[ \gamma_0 \otimes \gamma_0
\left(-K_0^3\Delta_k-U \right) \right.
\nonumber \\
&&\left.+a \vec \gamma \cdot \otimes \vec\gamma
\left( -K_0^3\Delta_k \right)
\right]
(2 \pi)^3
\delta^3(k) \ ,
\end{eqnarray}
where we dropped the sum in color and Dirac indices.
The factor $-3/4$
simplifies the color contribution for color singlets.
%fffffffffffffffffffffffffffffffffffffffffffffffffffffffffff
\begin{figure}
\begin{picture}(403,150)(0,0)
\put(-53,-26){\epsffile{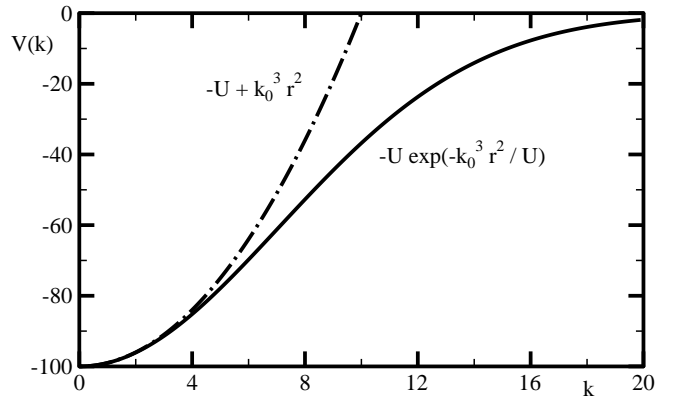}}\end{picture}
\caption{ $-U e^{- k_0^3 r^2 \over U}$ is an example of a potential
which tends to $-U +k_0^3 r^2$ in the limit of infinite U.
We illustrate this in the case where $K_0=1, \ U=100$.
\label{U finite}}
\end{figure}
%ffffffffffffffffffffffffffffffffffffffffffffffffffffffffffffffff
\par
The $\gamma_0 \cdot \gamma_0$ term in
the  potential is the limit of a series
of attractive potentials, see Fig. \ref{U finite},
and $U$ is an arbitrarily large infrared constant.
The infinite $U$ reappears in the self energy and
in color singlet channels this cancels the
infinitely attractive potential.
Any colored state will have a mass proportional to $U$ and will thus be
confined, see appendix A.
\par
The choice of an harmonic potential is not crucial, a linear
\cite{Adler} or funnel
potential has also been used, but a quadratic form is simpler.
In the case of light quarks the current quark mass $m_0$ is almost
vanishing and it essentially affects the family of the $\pi$ which is
a quasi Goldstone boson.
\par
The $\vec \gamma \cdot \vec \gamma$ term
is introduced in order to have Lorentz invariant pions
which are relativistic in the scalar decay.
Clearly the wrong result of a simple $\gamma_0 \otimes \gamma_0$
instantaneous interaction is the constant $f_\pi$ which is quite
small and is not Lorentz invariant.
The hope to cure $f_\pi$ with covariant extensions of the model
turned out to fail since they merely \cite{Kalinowsky} increased
$f_\pi$ in $30 \%$.
The value for the parameter $a$ which renders the pion Lorentz invariant
is $a=-.18$, and for most calculations (except for $f_\pi$ which is
increased by $ 300 \% $, see subsection D) this $a$ yields a result
comparable to the one we had for $a=0$.
The small $a$ suggests that the cure of $f_\pi$ may be related with
the pertubative short range \cite{Lagae} quark-quark interaction .
\par
Once $a$ is fixed, this model has the single scale of the
oscillator parameter $K_0$.
The simplest adimensional units of $K_0=1$ will
be used from now on in computations. When comparing with experiments
we will rescale $K_0$ to the value of $K_0=330 \, MeV$ which gives
the best overall fit of the meson spectrum.
We will also work in the momentum representation and drop the tilde
$\, \widetilde{ } \, $ from the potential.
%
%ssssssssssssssssssssssssssssssssssssssssssssssssssssssssssssssssssssssssss
\subsection{Chiral Symmetry Breaking at the BCS level with
quarks and antiquarks}
At the BCS level and with a color dependent interaction the
Schwinger Dyson equation for the quark self energy (which is
also the mass gap equation) is,
\begin{eqnarray}
\label{BCS MGE}
\displaystyle
\begin{picture}(20,5)
\put(20,0){\line(-1,0){20}}
\put(20,0){\vector(-1,0){10}}
\end{picture}^-1
\ &=& \ {\cal S}_0^{-1} -
\begin{picture}(30,10)
\put(5,0){\line(1,0){10}}
\put(25,0){\vector(-1,0){10}}
\put(5,0){$\cdot$}
\put(6,4.4){$\cdot$}
\put(8,7){$\cdot$}
\put(10.6,9){$\cdot$}
\put(15,10){$\cdot$}
\put(19.4,9){$\cdot$}
\put(22,7){$\cdot$}
\put(24,4.4){$\cdot$}
\put(25,0){$\cdot$}
\end{picture}
\nonumber \\
{\cal S}^{-1} &=& -i \not p - \Sigma
\end{eqnarray}
where the full (up to the approximation which is chosen)
propagator $\cal S$ is denoted as usual by
\begin{picture}(15,2)
\put(15,0){\line(-1,0){15}}
\put(15,0){\vector(-1,0){10}}
\end{picture} .
The subindex $_0$ is reserved for the free functions.
The effective quark-quark 2-body interaction of eq. (\ref{hamilt}),
a chiral invariant and color dependent interaction, is represented with a
dotted line $ \dots \dots$.
The Bethe Salpeter equation for the vertex is related by the Ward
Identities
with eq.(\ref{BCS MGE}), see Appendice C,
\begin{eqnarray}
\label{BCS BSE}
\displaystyle
\begin{picture}(15,5)
\put(5,2){\circle*{6}}
\put(5,3){\line(1,0){10}}
\put(5,1){\line(1,0){10}}
\end{picture}
\ &=& \ \Gamma_0 +
\begin{picture}(30,20)(0,10)
\put(20,10){\line(-2,1){20}}
\put(20,10){\vector(-2,1){12}}
\put(0,0){\line(2,1){20}}
\put(0,0){\vector(2,1){12}}
\multiput(0,-2)(0,2){11}{$\cdot$}
\put(20,10){\circle*{6}}
\put(20,11){\line(1,0){10}}
\put(20,9){\line(1,0){10}}
\end{picture}
\nonumber \\
\nonumber \\
\Gamma(p,q) &=& \Gamma_0 + \int {d^4k \over (2 \pi )^4}V(k)\Omega S(p+k)
\nonumber \\ &&
\Gamma(p+k,q+k) S(q+k) \Omega
\end{eqnarray}
where the full vertex $\Gamma^\mu$ is denoted by
\begin{picture}(15,4)
\put(2,2){\circle*{4}}
\put(2,3){\line(1,0){10}}
\put(2,1){\line(1,0){10}}
\end{picture}.
When eq.(\ref{BCS BSE}) is iterated, we find that it includes, the
Bethe-Salpeter ladder,
which will be represented diagrammatically by a box with 4 emerging lines
$ = \hspace{-.15cm} \sqcap \hspace{-.22cm} \sqcup \hspace{-.15cm} =$,
\FL
\begin{equation}
\label{ladder}
\begin{picture}(222,25)(0,0)
\put(0,0){\epsffile{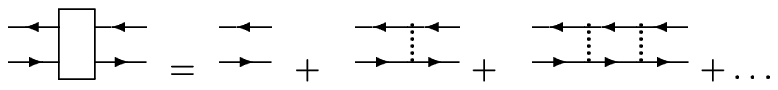}}\end{picture}
\end{equation}
where the ladder represents the mesons, see Appendice C.
In this way the quark propagator, the vertices and the
mesons are intertwined.
\par
In this case of an instantaneous interaction, it is convenient
to substitute the Dirac fermions in terms of Weyl fermions,
in order to find the hadron spectrum.
The Dirac propagator can be decomposed in a quark propagator and an
antiquark propagator, moving both forward in time.
\FL
\begin{eqnarray}
{\cal S}_{Dirac}(k_0,\vec{k})
&=& {i \over \not k -m +i \epsilon}
\nonumber \\
&=& {i \over k_0 -E +i \epsilon} \
{ 1+{m \over E} \beta + {k \over E} \vec \alpha \cdot \hat k \over 2}
\beta
\nonumber \\
&& - {i \over -k_0 -E +i \epsilon} \
{ 1-{m \over E} \beta - {k \over E} \vec \alpha \cdot \hat k \over 2}
\beta
\end{eqnarray}
It is convenient
to use \cite{papfpi,3P0} the quark energy projectors,
\begin{eqnarray}
\label{energy projectors}
 \Lambda^+=
{1 \over 2}\left(1+S\beta+C\widehat k \cdot \vec \alpha \right)
=\sum_su_su^{\dagger}_s \, ,
\nonumber \\
 \Lambda^-=
{1 \over 2}\left(1-S\beta-C\widehat k \cdot \vec \alpha \right)=
\sum_sv_sv^{\dagger}_s \, ,
\end{eqnarray}
where $S=\sin(\varphi)={m\over E} \ , \ C=\cos(\varphi)={k\over E}$
and $\varphi$ is a chiral
angle which in the non condensed case is equal to $\arctan{m_0 \over k}$,
($m_0$ is the current mass of the quark)
but is not determined from the onset when chiral symmetry breaking
occurs.
In this case the physical quark mass is a variational function
$m=m(k)$ which is determined by the mass gap equation. This is
equivalent to use the chiral angle $\varphi=\varphi(k)$ as the
variational function. In Fig. \ref{chiral} we show examples of non trivial
solutions for the function $\varphi(k)$.
\par
The energy projectors can be decomposed in the quark spinor
$u(k)$ and in the antiquark spinor $v(k)$, 
\begin{eqnarray}
u_s({\bf k})&=&{\Lambda^-\over\sqrt{ 1+S \over 2} }u_s(0)=\left[
\sqrt{ 1+S \over 2} + \sqrt{1-S \over 2} \widehat k \cdot \vec \alpha
\right]u_s(0)  \nonumber \\
v_s({\bf k})&=&
{\Lambda^-\over\sqrt{ 1+S \over 2} }v_s(0)=\left[
\sqrt{ 1+S \over 2} - \sqrt{1-S \over 2} \widehat k \cdot \vec \alpha
\right]v_s(0)  \nonumber \\
&=& -i \sigma_2 \gamma_5 u_s^*({\bf k}) \ .
\end{eqnarray}
And finally the Dirac quark propagator is decomposed in,
\begin{eqnarray}
\label{DiracWeyl}
{\cal S}_{Dirac}(w,\vec{k})
&=&u(k)
{\cal S}_{q}(p_0,\vec{k})
u^{\dagger}(k) \beta
\nonumber \\
&&- v^{\dagger}(k)
{\cal S}_{\bar q}(-p_0,-\vec{k})
v(k) \beta  \ ,
\end{eqnarray}
where the quark and antiquark Weyl propagators are.
\begin{equation}
{\cal S}_{q}(w,\vec{k})={\cal S}_{\bar q}(w,\vec{k})=
{i \over w-E(k) +i\epsilon} \ .
\end{equation}
The quark and antiquark formalism is convenient to calculate
the hadron spectroscopy.
With Weyl propagators the BS equation simplifies into the
Salpeter equation, in a form which is as close as possible to the more
intuitive Schr\"odinger equation.
In the Feynman rules with Weyl propagators, we
choose to redefine the vertices of the effective potential which now
include the spinors $u^\dagger \, , \ u \, , \ v^\dagger$ and $v$.
The ''$-$'' sign which affects the antiquark propagator
in eq.(\ref{DiracWeyl}) could also be included in the vertices with
$v^\dagger$, but we prefer to recover the equivalent rules which are common
to nonrelativistic field theory. This ''$-$'' sign together with the one
from the fermion loops will be included in the antiquark vertex
and in diagrams with quark exchange or with antiquark exchange.
The Dirac vertex $\gamma_0$ is now replaced by
$u^{\dagger}u \, , u^{\dagger}v \, , v^{\dagger}u $ or $-v^{\dagger}v$
when the vertex is respectively connected to a quark,
a pair creation, a pair annihilation or an antiquark; and the Dirac
vertex $\vec \gamma$ is respectively replaced by
$u^{\dagger}\vec \alpha u \, , u^{\dagger}\vec \alpha v \,
, v^{\dagger}\vec \alpha u $ or $-v^{\dagger}\vec \alpha v$.
We choose the graphical notation for the Weyl propagators of quarks
and antiquarks,
\begin{eqnarray}
{\cal S}_{Dirac}(w,\vec{k}) &=&
\begin{picture}(25,15)(0,0)
\put(15,6){\vector(-1,0){10}}
\put(5,6){\line(-1,0){5}}
\put(5,15){$_{w,\vec{k}}$}
\put(20,0){$D$}
\end{picture}
\ \ \ , \nonumber \\
{\cal S}_{q}(w,\vec{k}) =
\begin{picture}(15,15)(0,0)
\put(15,1){\vector(-1,0){10}}
\put(5,1){\line(-1,0){5}}
\put(5,10){$_{w,\vec{k}}$}
\end{picture}
\ \ \ &,& \
{\cal S}_{\bar q}(-w,-\vec{k}) =
\begin{picture}(15,15)(0,0)
\put(0,1){\vector(1,0){10}}
\put(10,1){\line(1,0){5}}
\put(5,10){$_{w,\vec{k}}$}
\end{picture}
\end{eqnarray}
where the Diagrams using the Feynman rules corresponding to the
Dirac fermion propagators will have a subindex $_D$ in the
remaining of the paper. In the
case of the Weyl propagators (which will be used more often
the the Dirac propagators)
the quark will be represented with an arrow pointing to the
left while the arrow pointing to the right represents an
antiquark (both move forward in the time direction).
%
%sssssssssssssssssssssssssssssssssssssssssssssssssssssssssssssssssssssssssss
\subsection{The BCS mass gap equation and the quark energy}
\begin{figure}
\begin{picture}(403,150)(0,0)
\put(-58,-26){\epsffile{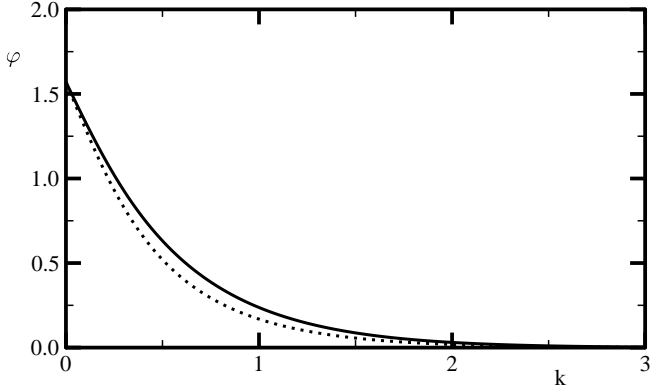}}\end{picture}
\caption{ We show the BCS chiral angle in units of $K_0=1.$
We also represent with a doted line the chiral angle
that we obtain going beyond BCS.
\label{chiral}}
\end{figure}

Here we derive the mass gap equation, and the quark dispersion relation,
replacing the propagator of eq. \ref{DiracWeyl} in the Schwinger Dyson
equation for the quark  self energy (\ref{BCS MGE}), 
\begin{equation}
\label{eqnova}
u \
\begin{picture}(15,15)(0,0)
\put(15,1){\vector(-1,0){10}}
\put(5,1){\line(-1,0){5}}
\put(3,-5){$_{w,\vec{k}}$}
\end{picture}
{ }^{-1}
u^\dagger
-
v \
\begin{picture}(15,15)(0,0)
\put(0,1){\vector(1,0){10}}
\put(10,1){\line(1,0){5}}
\put(3,-5){$_{w,\vec{k}}$}
\end{picture}
{ }^{-1}
v^\dagger
=
\beta
{\not k -m_0 \over i}
-
\beta
\begin{picture}(30,10)
\put(5,0){\line(1,0){10}}
\put(25,0){\vector(-1,0){10}}
\put(5,0){$\cdot$}
\put(6,4.4){$\cdot$}
\put(8,7){$\cdot$}
\put(10.6,9){$\cdot$}
\put(15,10){$\cdot$}
\put(19.4,9){$\cdot$}
\put(22,7){$\cdot$}
\put(24,4.4){$\cdot$}
\put(25,0){$\cdot$}
\put(27,-7){$D$}
\end{picture}
\end{equation}
Another equivalent method is to use the Hamiltonian formalism
for the quark and antiquark creators and annihilators \cite{3P0},
and find the Bogoliubov Valatin transformation which would
minimize the vacuum energy density. In that Hamiltonian formalism
the mass gap equation is also obtained when the quark
antiquark pair creation operators are postulated to vanish in
the Hamiltonian, in order to ensure the vacuum stability against
spontaneous generation of scalars. With the present method
we project eq(\ref{eqnova}) with the spinors $u^\dagger \cdots u$
and $u^\dagger \cdots v$, and we get directly the quark and antiquark
energy and the mass gap equation,
\begin{eqnarray}
\label{2 eqs}
&&E(k) = u_s^\dagger(k) \left\{k \widehat k \cdot \vec \alpha + m_0 \beta
-\int {d w' \over 2 \pi} {d^3k' \over (2\pi)^3}
i V_l(k-k')  \right.
\nonumber \\
&&\left. \left[ {\Omega_l \Lambda^+(k') \Omega_l \over w'-E(k') +i\epsilon}
-{  \Omega_l \Lambda^-(k') \Omega_l \over -w'-E(k')+i\epsilon} \right]
\right\} u_s(k),
\nonumber \\
&&0 = u_s^\dagger(k) \left\{k \widehat k \cdot \vec \alpha + m_0 \beta
-\int {d w' \over 2 \pi} {d^3k' \over (2\pi)^3}
i V(k-k') \right.
\nonumber \\
&&\left. \left[ {\Omega_l \Lambda^+(k') \Omega_l \over w'-E(k') +i\epsilon}
-{\Omega_l \Lambda^-(k') \Omega_l  \over -w'-E(k')+i\epsilon} \right]
\right\} v_{s''}(k) \, .
\end{eqnarray}
In the case of an instantaneous interaction, the loop integral in the energy
$w$ removes the pole in the propagator,
\begin{equation}
\int {d w \over 2 \pi} {i \over w -E(k) + i \epsilon} =
{ 1 \over 2}
\end{equation}
and in the case of a quadratic interaction, the loop
integral in the momentum is transformed in a Laplacian,
see eq.(\ref{interaction}).
Some useful properties are,
\FL
\begin{eqnarray}
\label{spinor properties}
u^\dagger_s  u_{s'} = 1 \delta_{ss'}
\ \ &,& \ \
u^\dagger_s  v_{s'} =
 0 \left[\vec \sigma \cdot \hat k i \sigma_2\right]_{ss'}
\nonumber \\
u^\dagger_s  \beta u_{s'} = S \delta_{ss'}
\ \ &,& \ \
u^\dagger_s  \beta v_{s'} =
- C \left[\vec \sigma \cdot \hat k i \sigma_2\right]_{ss'}
          \\
u^\dagger_s  \vec \alpha \cdot \hat k u_{s'} =
C \delta_{ss'}
\ \ &,& \ \
u^\dagger_s  \vec \alpha \cdot \hat k v_{s'} =
S \left[\vec \sigma \cdot \hat k i \sigma_2\right]_{ss'}  
\nonumber \\
u^\dagger_s  \beta \vec \alpha \cdot \hat k u_{s'} =
0 \delta_{ss'}
\ \ &,& \ \
u^\dagger_s  \beta \vec \alpha \cdot \hat k v_{s'} =
1 \left[\vec \sigma \cdot \hat k i \sigma_2\right]_{ss'}   \, .
\nonumber
\end{eqnarray}
We get finally for the quark energy,
\begin{eqnarray}
\label{final 2 eqs}
E(k) &=& kC +m_0S +{1 \over 2} \left[
+S \Delta (S) +C \hat k \cdot \Delta ( \hat k C)  \right] +{U \over 2}
\nonumber \\
&& +a {1 \over 2} \left[
-3 S\Delta (S) -C \hat k \cdot \Delta ( \hat k C)  \right]
\nonumber \\
&=& {U \over 2} +k C +m_0S -{ {\dot \varphi }^2 \over 2 }
- { C^2 \over k ^2 }
\nonumber \\
&&-a \left[ SC \Delta\varphi -{C^2 \over k^2}
-\left( S^2+{1 \over 2}  \right) {\dot \varphi}^2\right] \, ,
\end{eqnarray}
where in color singlets the $U/2$ term will cancel the $-U$ term
from the 2-body quark potential.
For the mass gap equation we get,
\begin{eqnarray}
\label{final 2 mgeqs}
0 &=& \Bigl\{kS -m_0C +{1 \over 2} \left[
-C \Delta (S) +S \hat k \cdot \Delta ( \hat k C)  \right] \Bigr.
\nonumber \\
&& \Bigl. +a {1 \over 2} \left[
+3C \Delta (S) -S \hat k \cdot \Delta ( \hat k C)  \right] \Bigr\}
\left[\vec \sigma \cdot \hat k i \sigma_2\right]_{ss'} 
 \nonumber \\
 &=&-\Delta \varphi  +2kS -2m_0C -{2SC \over k^2}
\nonumber \\
 &&-a \left[ -\left(2 C^2 +1 \right) \Delta \varphi 
+ 2 SC \left( {\dot \varphi}^2-{1 \over k^2} \right) \right]\, .
\end{eqnarray}
The mass gap equation is in general a nonlinear integral equation, but
in this case of a harmonic potential it simplifies to a differential
equation. We solve it numerically with the Runge-Kutta and shooting
method, see Fig. \ref{chiral} for the solution.
%ffffffffffffffffffffffffffffffffffffffffffffffffffffffffffff
%ffffffffffffffffffffffffffffffffffffffffffffffffffffffff
%
%sssssssssssssssssssssssssssssssssssssssssssssssssssssssssssssssssssssssss
%
\subsection{The pseudoscalar and scalar solutions
 to the Salpeter equation}
The homogeneous Salpeter equation for a meson (a color singlet
quark antiquark boundstate) is, according to Appendix C,
\FL
\begin{eqnarray}
\label{homo sal}
&+M(P)-E(k_1)-E(k_2)
\over i
&\phi^+(k,P)=-iu^\dagger(k_1)
\chi(k,P)  v(k_2)
\nonumber \\
&-M(P)-E(k_1)-E(k_2)
\over i
&{\phi^-}^t(k,P)=-iv^\dagger(k_1)
\chi(k,P) u(k_2)
\nonumber \\
&\chi(k,P)=
\int {d^3k' \over (2 \pi)^3} &V_l(k-k') \Omega_l \left(
u(k'_1)\phi^+(k',P)v^\dagger(k'_2)
\right.
\nonumber \\
&&\left.
+v(k'_1){\phi^-}^t(k',P) u^\dagger(k'_2)\right) \Omega_l
\end{eqnarray}
where $k_1=k+{P \over 2} \ , \ k_2=k-{P \over 2}$ and $P$ is
the total momentum of the meson.
We use the Bethe-Salpeter amplitude $\chi$ as an intermediate step
to compute the contribution of interaction $V$ to the boundstate equation.
The wave functions $\phi^+$ and $\phi^-$ are equivalent
to the Bethe-Salpeter amplitude $\chi $ .
For color singlets the contribution of the infinite infrared
constant $U$ cancels, see Appendix A.
The equation is also
flavor independent, and we will now concentrate on the
$momentum \, \otimes \, spin$ part of the wave-functions.
We will now drop the $U$ term and the color dependence from
the equations. In this section the matrices $\Omega_l$ will
only include the Dirac structure,
$\Omega_l \otimes \Omega_l=\gamma_0 \otimes \gamma_0
+a \vec \gamma \cdot \otimes \vec\gamma$.
With the aim of studying the $f_0$ decay in a pair of $\pi$,
we will now solve the bound state equation
equation for the scalar $f_0$ in its center of mass frame and
the equation for the pseudoscalar groundstate $\pi$ in the limit of
small $P$ and in the limit of large $P$.
\par
Due to the large mass of the scalar meson $f_0$ in this model,
it turns out that the negative energy $\phi^-$ component for
the groundstate is less than $10\%$ of the positive energy
component $\phi^+$.
The Schr\"odinger limit, where only the positive energy component is
considered, is therefore acceptable.
A general form for the $^3P_0$ wave-function for the scalar is
\begin{equation}
\label{scalar wf gen}
\phi^+(k)_{s_1s_2}=
k \phi_s(k){ [\vec \sigma \cdot \hat k i \sigma_2]_{{s_1}{s_2}}
\over \sqrt{2}}
\end{equation}
the truncated BS amplitude is then,
\FL
\begin{eqnarray}
\chi_s&=&- \Omega_l \Delta {2 k \phi_s(k) \over 1+S}
\Lambda^+
u_{s_1}(0) {[\vec \sigma \cdot \hat k i \sigma_2]_{{s_1}{s_2}}
\over \sqrt{2}}
v^{\dagger}_{{s_2}}(0)
\Lambda^-  \Omega_l \nonumber \\
&=& \Omega_l \Delta {\Lambda^+ \beta \vec \alpha \cdot \vec k \phi_s(k)
\over \sqrt{2}} \Omega_l  \ ,
\end{eqnarray}
we get for left hand side of eq. (\ref{homo sal}) ,
\begin{eqnarray}
u^\dagger(k) \chi(k,0) v(k) ={1+a \over 2} \Delta \left(\phi^+\right)
&&+{1-a \over 2} S\Delta \left(S\phi^+\right)
\nonumber \\
+{1-3a \over 2} C\Delta \left(C\phi^+\right)&&+{1-3a \over k^2}C^2 \phi^+
\end{eqnarray}
and the radial Salpeter equation for the scalar in the center of mass is,
\FL
\begin{eqnarray}
&&\left[ 2E(k)-M -\left({d^2 \over dk^2} -{2\over k^2}
- {{\dot \varphi}^2 \over 2}+ {C^2 \over k^2}
\right)
-a \left( -C^2 {d^2 \over dk^2} \right. \right.
\nonumber \\
&&\left. \left.
+2SC\dot \varphi {d \over dk}
-\hspace{-.01cm}{C^2 \over k^2} +\hspace{-.01cm}SC\ddot \varphi
+{1 +2 C^2\over 2}{\dot \varphi}^2 \right) \right]k^2\phi_s =0
\end{eqnarray}
Solving the bound state equation we find that
the solution of the equation is very close to a Gaussian,
\begin{eqnarray}
\label{scalar wf}
\phi_s(k)  \simeq  { e^{-k^2 \over 2 \alpha_s^2} \over {\cal N}_s}
\ , \
{\cal N}_s^{-1} = { 4 \sqrt{ \pi \sqrt \pi} \over \sqrt 3 \alpha_s^{5/2} }
\ , \
\alpha_s \simeq .476
\end{eqnarray}
and the mass is $M=2.94\,K_0=970\,MeV$ which is close to the
most probable experimental mass of the $f_0$ groundstate.
\par
We now study the pseudoscalar groundstate in the low $P$ limit,
which was already studied extensively in the literature
\cite{papfpi,ourcouca}.
For vanishing $P$ we find that $\phi^+=-\phi^-$ and both are
proportional to $\sin(\varphi)$. This is due to the Goldstone boson
nature of the $\pi$, see the result of Appendix C.
However this component of the wave-function has zero norm, and it
is necessary to include the next order of the expansion in $P$
to determine the norm.
The most general low $P$ pseudoscalar wave-function is then,
\FL
\begin{eqnarray}
\label{pip0}
&&\phi^+={\cal N}_p^{-1}\left(S + {M(P) \over k} f_1 + i
g_1 { \vec P \over k} \cdot \hat k \times \vec \sigma \right)
{i \sigma_2 \over \sqrt{2}} \nonumber \\
&&\phi^-={\cal N}_p^{-1}\left(-S + {M(P) \over k} f_1 - i
g_1 {\vec P \over k} \cdot \hat k \times \vec \sigma \right)
{i \sigma_2 \over \sqrt{2}} \, ,
\end{eqnarray}
where the norm is a function of the $\pi$ energy,
\begin{eqnarray}
{\cal N}_p^2&=& 
\bigl({2 f_{\pi}^{(t)}}\bigr)^2 \, M  \ \ , \ \ \
M^2(P)=M^2(0)+P^2\sqrt{f_{\pi}^{(s)} \over f_{\pi}^{(t)}}
\nonumber \\
M^2(0)&=&-{2m_0 \langle \bar \psi \, \psi \rangle \over {f_\pi^{(t)}}^2}
\ \ , \ \ \
\langle \bar \psi \, \psi \rangle = - 6 \int { d^3k \over (2 \pi)^3} S
\end{eqnarray}
and where in the case of an instantaneous interaction there are
\cite{papfpi,ourcouca} usually 2 different
$f_{\pi}^{(t)}$ and $f_{\pi}^{(s)}$,
\begin{eqnarray}
f_\pi^{(t)}&=& \sqrt{ {3 \over \pi ^2 }
\int_0^\infty d k \ k f_1 S}
\\ \nonumber
\sqrt{ f_\pi^{(s)} f_\pi^{(t)} }&=& \sqrt{ {1 \over 2 \pi ^2 }
\int_0^\infty d k \ -k^2S \dot \varphi + 4 k C (g_1+1/2) }  \, .
\end{eqnarray}
Substituting the wave-functions of eq. (\ref{pip0}) in
eq. (\ref{homo sal}), and expanding the resulting equation
up to the first order in $P$, we get the equations for the $f_1$
and $g_1$ components,
\begin{eqnarray}
\ddot f_1 \
&=&{1 \over 1 +a(2S^2+1)}
\Bigl[ -kS +  (2Ck) f_1 \Bigr.
\nonumber \\ &&
+4a\left(-SC \Delta \varphi -C^2 / k^2
+S^2{\dot \varphi}^2 + {\dot \varphi}^2/2\right)f_1
\nonumber \\ &&
\Bigl. -4aSC\dot \varphi (\dot f_1 - f_1 / k) \Bigr]
\label{f1} \, ,
\\
\ddot g_1 \
&=&{1 \over 1 +a(2S^2-1)}
\Bigl[ k C +  (2k C+2S^2/k^2) g_1 \Bigr.
\nonumber \\
&&
+2a(S^2/k^2 -2SC\Delta \varphi -2 S^2 {\dot \varphi}^2 )g_1
\nonumber \\
 &&
\Bigl. -4aSC\dot \varphi (\dot g_1 - g_1 / k) \Bigr]
\label{g1} \, .
\end{eqnarray}
It turns out that the parameter $a$ has little effect on most
functions, except for $f_1$.
The homogeneous equation for $f_1$ has the solution
$a_0=-.195$ an thus $f_1 \alpha 1/a-a_0$.
This will essentially affect $f_\pi^{(t)}$, $f_\pi^{(s)}$ and
the pion velocity $c$.
We find for $a \simeq -.18$ that $c=1$,
$f_\pi^{(t)}=f_\pi^{(s)}=0.21K_0\simeq 69 \, MeV$.
This shows a clear improvement of the model, with a correct relativistic
pion and a better $f_\pi$.
%ffffffffffffffffffffffffffffffffffffffffffffffffffffffffffff
\begin{figure}
\begin{picture}(403,150)(0,0)
\put(-58,-26){\epsffile{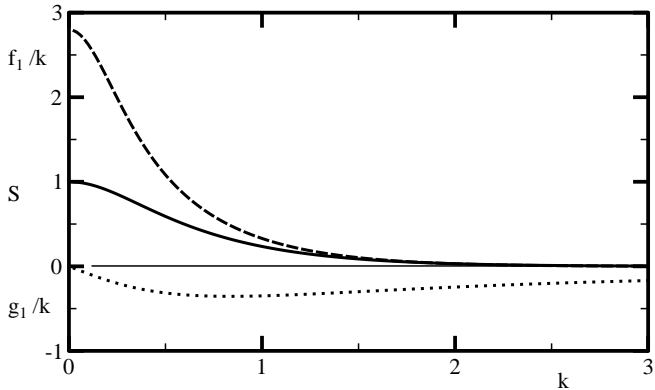}}\end{picture}
\caption{ We represent the $\pi$ wave-functions functions
$S$, $f_1/k$ and $g_1/k$ respectively with solid, dashed and
dotted lines, in the adimensional units $K_0=1$.
\label{sinf1g1}}
\end{figure}
%ffffffffffffffffffffffffffffffffffffffffffffffffffffffff
\par
We now discuss the pseudoscalar groundstate in the other limit of large
momentum $P$.
In this case the negative energy
components are suppressed by a factor of $1/P$.
The chiral angle $\varphi$, depicted in
Fig. [\ref{chiral}] vanishes completely, and the
spinors are simpler, for instance,
\begin{equation}
u_{s}(k_1) \simeq
{ 1+ \vec \alpha \cdot \hat k_1
\over \sqrt{2} }u_s(0) \ , \,  \hat k_1 \simeq \hat P +
{2 \over P} \vec k _\bot
\end{equation}
where the index $_\bot$ denotes the projection
$\vec k - (\vec k \cdot \widehat P ) \widehat P$
of a vector $\vec k$ in the plane perpendicular to $\vec P$.
The vertices, up to first order in $1/P$ are for instance,
\FL
\begin{eqnarray}
\label{large P}
u^\dagger_{s}(k_1)u_{s'}(k'_1)&\simeq&
\delta_{ss'} -{1 \over P}i\vec \sigma _{ss'} \hat P \times
(\vec k_\bot -\vec k'_\bot)
\\
u^\dagger_{s}(k_1) \vec \alpha u_{s'}(k'_1)
&\simeq& \hat P \delta_{ss'}+{
(\vec k_\bot +\vec k'_\bot) +i \vec \sigma \times
(\vec k_\bot -\vec k'_\bot) \over P}
\nonumber
\end{eqnarray}
Up to first order in $2 / P$ the equation for positive and negative
energy boundstate functions $ \phi^\pm(k) i \sigma_2 / _{\sqrt{2}} \, $
is,
\FL
\begin{eqnarray}
\label{posi}
0&\simeq &\left(P +{2 k^2 \over P} \mp M -(1-a)\Delta_{k} \right)
\phi^\pm(k)
\nonumber \\
&&-2 i(1-a){\widehat P \over P} \times \vec \nabla_k \cdot \left\{
\vec \sigma , \phi^\pm(k) \right\}
\nonumber \\
&&-a \Delta_k \phi^\mp(k)
-{2 \over P^2}\vec \sigma_\bot \cdot \phi^\mp(k)  \vec \sigma_\bot
\end{eqnarray}
We find that the wave-function has a component
with structure $i \vec \sigma \cdot \widehat P \times \vec k$.
However this component is smaller than the s-wave component by a factor
of less than $1/P$.
Thus we find that the $momentum \, \otimes \, spin$
solution up to highest order in, is essentially a
positive energy Gaussian function
$\phi_p(k) { i \sigma_2 \over \sqrt 2}$ ,
\begin{eqnarray}
\displaystyle
\label{large P wf}
\phi_p(k) \simeq  { e^{-k^2 \over 2 \alpha_p^2} \over {\cal N}_p}
, \
{\cal N}_p^{-1} = \left( { 2 \sqrt{ \pi } \over \alpha_p } \right)^{3\over2}
, \
\alpha_p^2 = \sqrt{{1-a \over 2}P}
\end{eqnarray}
For large momentum $P$ we find that $\phi^+$ is quite flat in k, while
$\phi^-$ is almost negligible,
\begin{equation}
\phi^-\simeq{a \over 2p} \Delta \phi^+ \ .
\end{equation}
This result is consistent with the
relativistic space contraction. We checked that the components
that we neglected here would yield a small contribution to the $f_0$
decay.
%
%sssssssssssssssssssssssssssssssssssssssssssssssssssssssssssssssssssssssss
\subsection{The coupling of a scalar to a pair of pseudoscalars}
The form factor $F(P)$ for the coupling of a scalar $f_0$
to a pair of $\pi$ can be decomposed in diagrams where 
a quark (antiquark) line either emits (absorbs) a pseudoscalar 
or a scalar. We use the
truncated Bethe-Salpeter amplitude $\chi$, as an intermediate step
to compute the coupling of a meson to a quark line $u^\dagger \chi u$.
$F(P)$ is represented with a large triangle and the boundstate amplitudes
$\chi$ and $\phi$ are represented with small triangles,
\FL
\begin{eqnarray}
\label{scalar coupling}
\begin{picture}(220,60)(0,0)
\put(0,0){\epsffile{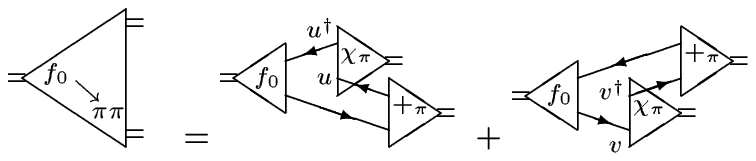}}\end{picture}
\nonumber \\
\begin{picture}(220,50)(0,0)
\put(0,0){\epsffile{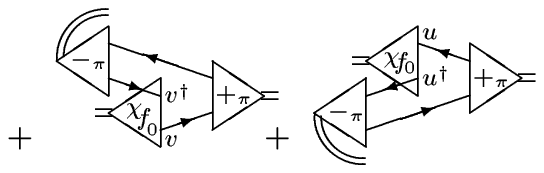}}\end{picture}
\end{eqnarray}
This includes the $q \bar q$ pair creation or annihilation of Fig.
\ref{quartic diagrams}. The same irreducible interaction for quarks 
which is used in the boundstate equations is also \cite{Hecht} used 
for the annihilation.
In eq. (\ref{scalar coupling}) the loop energies are trivial, see 
for instance eq. (\ref{annihilation overlaps}),
and we now compute the $momentum \otimes spin$ contribution.
\par
We first consider the limit of a coupling to pions of low momentum $P$.
The coupling of the pseudoscalar to the quark is derivative and thus
is suppressed.
For instance in the case of a massless $\pi$ in the center of mass,
using the wave-function of eq. (\ref{pip0}), we find that
\FL
\begin{eqnarray}
&& \chi= \left[ -(1-3a)\sqrt{2} {\cal N}_p^{-1} \,
\Delta \left(S\right) \right] \beta \gamma_5 +o(P)
\nonumber \\
&& \Rightarrow u^\dagger(k) \chi(k,0) u(k)
=o(P)  \ ,
\end{eqnarray}
which is consistent with the derivative coupling of a pion to
a quark.
The dominant contribution includes the coupling of the $^3P_0$
scalar meson to the quark (antiquark) line.
The coupling $F(P)$ to a pair of pions with low momentum $P$ is,
\FL
\begin{eqnarray}
\label{low F}
F &=&  tr \int {d^3 k \over (2 \pi)^3}
{\phi^-}^\dagger \left(  u^{\dagger}\chi_s u
\phi^+ +\phi^+ v^{\dagger}\chi_s v  \right)
\end{eqnarray}
where for instance we get for the scalar coupling
$u^{\dagger}(k)\chi_s(k)u(k)$ to the quark line,
\begin{eqnarray}
u_{s_1}^{\dagger}\chi_s(k)u_{{s_2}} &= {
\delta_{s_1s_2} \over 2 \sqrt 2 }& \Bigl[
(1-a)C \hat k \cdot \Delta ( \hat k S k \phi_s )
\Bigr.
\nonumber \\
&& \Bigl.
- (1-3a)S \Delta ( C k \phi_s ) \Bigr]
\end{eqnarray}
except for the $-i$ factor which goes with any potential
insertion according to the Feynman rules. We will discard
it in this section.
Integrating by parts the eq.(\ref{low F}) can be simplified with the
help of the mass gap equation and we get,
\FL
\begin{eqnarray}
F&=& \int {d^3 k \over (2 \pi)^3}
{k \phi_s \over \sqrt{2} }
\left[-kS +\dot \varphi {d \over dk} +a SC \Delta \right]
\left( {\phi^-}^\dagger \phi^+ \right)
\nonumber \\
&=& {0.011{K_0}^2-0.052 {M^{}}^2(P) \over f_\pi^2 M(P)} K_0^{1/2}  \ .
\end{eqnarray}
This coupling is very sensitive to the pion decay constant $f_\pi$
and to the energy $M(P)$ of the pion.
\par
We now consider the opposite limit of large pion momentum $P$.
In this case the negative energy component of the pion is quite small.
We expect that the dominant diagrams are the ones of the
first line of eq. (\ref{large P wf}), which include only the positive
energy $\phi^+$. This is present
in either the coupling of a pion to the quark line,
or the coupling of a $\pi$ to an antiquark line,
\FL
\begin{eqnarray}
F(P)&=&\int {d^3 k \over (2 \pi)^3}
\phi_s(k_1,0) (-\Delta{_k'}) \Bigl[
\phi_p(k',P) \phi_p(k,-P)t_q
\nonumber \\ &&
+\phi_p(k,P) \phi_p(k',-P)t_{\bar q}\Bigr] \Biggr|_{k'=k} \, ,
\end{eqnarray}
where $t_q$and $t_{\bar q}$ are respectively the traces
in Dirac indices,
\FL
\begin{eqnarray}
t_q &=&tr \Bigg\{ {-i \sigma_2 \over \sqrt 2}\vec \sigma \cdot \vec k_1
 u^\dagger(k_1)\Omega_l u(k'_1){i \sigma_2 \over \sqrt 2 }v^\dagger(k'_2)
\Omega_l u(k_2) {i \sigma_2 \over \sqrt 2 }  \Bigg\}
\nonumber \\
&=& tr \Bigg\{ {\vec \sigma \cdot \vec k_1  \over \sqrt 2 }
{(1+\beta) \over 2}
{(1+\vec \alpha \cdot\hat k_1 ) \over \sqrt 2 } \Omega_l
{(1+\vec \alpha \cdot\hat k'_1 ) \over \sqrt 2 }
\nonumber \\ &&
{1 \over \sqrt 2 }
{(1+\beta) \over 2}
\gamma_5
{(1-\vec \alpha \cdot\hat k'_2 ) \over \sqrt 2 } \Omega_l
{(1+\vec \alpha \cdot\hat k_2 ) \over \sqrt 2}
{1 \over \sqrt 2 } \Bigg\}
\end{eqnarray}
and $t_{\bar q}$, of the second diagram , with the coupling
of the pion to the antiquark yields the same result.
The total coupling is a functional of the scalar and
pseudoscalar wavefunctions, which are described
in eqs. (\ref{scalar wf},\ref{large P wf}) by Gaussians.
We now apply the Laplacian to the functions of $k'$.
The dominant term of the expansion in $2/P$ comes from the derivatives
of $\phi_P$.
A derivative of $\phi_P(k',P)$ will be proportional to $ P / 2$ when the
Gaussian integral is performed , while a derivative of
$t_q$ or $t_{\bar q}$ which are functions of
respectively $\hat k'_1$ or $ \hat k'_2$ is proportional to $1/P$
and will not produce a dominant term.
The traces then simplify to,
\begin{eqnarray}
\label{traces}
t_q=t_{\bar q}= a \,
{k_1 \over \sqrt 2 } \left[1 - (\hat k_1 \cdot \hat k_2)^2\right]
\end{eqnarray}
We now apply the Laplacian to $\phi_P$, and expand the
$\hat k_1 \cdot \hat k_2$ in a series of $1/P$.
It is convenient to define $\alpha_T^2= 2 \alpha_s^2 + \alpha_p^2$,
and we get finally find for the  $momentum \otimes spin$
contribution,
\FL
\begin{eqnarray}
\label{scalar f(P)}
\displaystyle
F(P) &=& \left[a {- 64 \pi^{3 \over 4} \ \ \alpha_s^{1\over2} \alpha_p^2
\over 3^{1\over2} \,
\left(\alpha_p^2+\alpha_T^2\right)^2 \alpha_T^3}{P\over 2}
+ o\left({2\over P}\right) \right]
e^{-{\left({P\over 2}\right)^2 \over \alpha_T^2}}
 \end{eqnarray}
\par
where the dominant term is the $a$ term which is of the order
of $0.43 \, K_0^{-1 / 2}$ for $P$ of the order of $2 K_0$.
\par
The color factor for$f_0$ and $\pi$ color singlets is $1 / \sqrt{3}$.
The flavor factor for the coupling of a scalar isosinglet
$(u \bar u + d \bar d) / \sqrt{2}$ to a pair of pseudoscalar
isovectors, say $ u \bar d$ and  $-d \bar u$, with a flavor
independent quark-antiquark annihilation is $-1 / \sqrt{2}$.
The total coupling is then $ - F(P) / \sqrt{6}$.
\par
\begin{figure}
\begin{picture}(403,155)(0,0)
\put(-53,-26){\epsffile{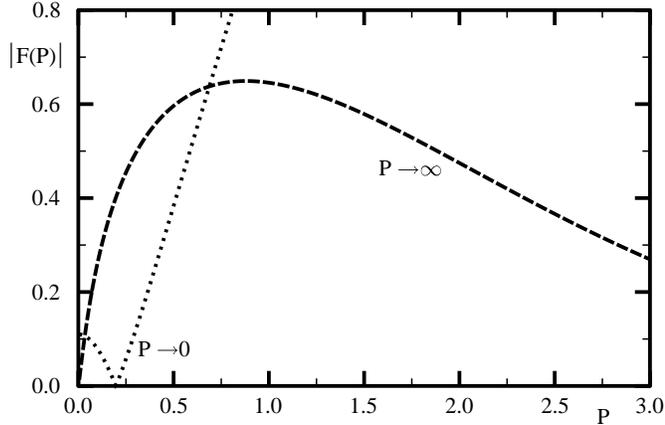}}\end{picture}
\caption{
We show $F(P)$, in the adimensional units of $K_0=1$.
The dotted line and dashed line correspond respectively low $P$
and high $P$ limits.
\label{F(P) 2 limits}}
\end{figure}

The function $F(P)$ is very cumbersome to derive in the case of
intermediate momenta. For momenta of the order of $K_0$, see
Fig. \ref{F(P) 2 limits}, matching the high and low $P$ limits
with an interpolating function is a possible approximation.
%ffffffffffffffffffffffffffffffffffffffffffffffffffffffffffffffff
%fffffffffffffffffffffffffffffffffffffffffffffffffffffffffffffffff
%
%sssssssssssssssssssssssssssssssssssssssssssssssssssssssss
\subsection{The $ f_0(980) \rightarrow \pi \pi$  decay}
The decay width of a $f_0$ in a pair of $\pi$ can be calculated
from the the Breit-Wigner pole in the meson propagator. We call
bare the ladder meson, see eqs. (\ref{ladder}), (\ref{4 S matrices})
and (\ref{last ladder}).
The bare mass $M_0$ is real and is a solution of the
Bethe-Salpeter equation.
When coupled channels of mesons are included, the bare meson is
dressed. The dressed pole
is composed by the bare mass $M_0$ plus the coupled channel
contribution which includes a real mass shift, and an
imaginary term in the case where the mass is above the coupled
thresholds. The mass is then,
\FL
\begin{eqnarray}
&& \ \ \ \ M=M_0 + \Delta M \ , \ \ \ \Delta M = -i \Sigma
\nonumber \\
&& \label{scalar self energy}
\begin{picture}(200,50)(0,0)
\put(0,0){\epsffile{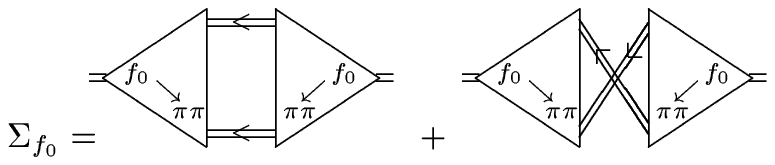}}\end{picture}
\end{eqnarray}
where we only included a loop of bare $\pi$, which is the
simplest contribution to the scalar self energy.
The integral in the loop energy provides an extra $2 \pi i$ factor,
which verifies that in general $\Delta M$ includes
a real component, and we  find that,
\FL
\begin{equation}
\Delta M = 6
\int {d^3q \over (2 \pi)^3} { {F(q)^* \over \sqrt{6}}
\ {F(q) \over \sqrt{6}}
\over M_{f_0} - 2 \sqrt{q^2 +M_{\pi}^2} +i\epsilon } 
\end{equation}
where the factor of $6$ includes the 3 different
flavors of the isovector $\pi$, and the factor 2 from the
direct and exchange diagram of the self energy in eq.
(\ref{scalar self energy}). In this
section we represent as usually the width of a resonance by a
$\Gamma$, which should not to be confused with the same symbol which
is used in other sections for the vector or axial vertices.
The width $\Gamma$ is a simple function of the imaginary component
of $\Delta M$,
\begin{eqnarray}
\label{decay}
Im( \Delta M) = -i {\Gamma \over 2}  \ , \  \
\Gamma = {1\over 4 \pi} P \, M_{f_0} \, |F(P)^2|
\end{eqnarray}
where
$P = { 1 \over 2} \sqrt{M_{f_0}^2-4M_{\pi}^2}$
is the momentum of the emitted $\pi$.
Let us consider the case of a scalar mass $M_{f_0}$ of the order of
$1 \, GeV$, where $P$ is larger than the scale
of the interaction by a factor of $1.4$. In this case, it is
sensible to use the limit of large momentum $P$ for $F(P)$,
see eq. (\ref{scalar f(P)}).
We finally find a partial decay
width of just $40 \, MeV$ for $F_0 \rightarrow \pi \pi$ which lies
within the experimental limits. We expect that a complete calculation
without the large $P$ approximation would not deviate
from this by more than a factor of $2$.
\par
This is also compatible with the narrow resonances $f_0(1500)$ and 
$a_0(980)$ which are possible groundstates. Concerning $K_0^*(1430)$ 
which is wider and has decay products with a larger momentum $P$, 
the function $\Gamma(P)$ of eq.(\ref{decay}) has the correct 
qualitative behavior of being proportional to $P^3$ for intermediate
momenta. However in this model the exponential decrease is too strong,
and the model needs some improvement in order to reproduce the
correct $K_0^*(1430)$.
%
%sssssssssssssssssssssssssssssssssssssssssssssssssssssssssssssssssssssssss
%
\section{Going Beyond BCS with Finite Coupled Channel Effects}
%
%sssssssssssssssssssssssssssssssssssssssssssssssssssssss
%
\subsection{The Mass Gap Equation and the self energy}
\par
We find in Appendix D that the minimal extension of the mass
gap equation beyond BCS is achieved with a new tadpole term
in the self energy,
\begin{eqnarray}
\label{CS3 self}
\begin{picture}(110,40)(0,0)
\put(0,0){\epsffile{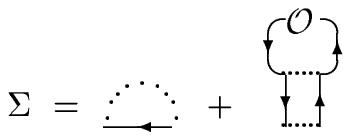}}\end{picture} \ ,
\nonumber \\
\begin{picture}(150,50)(0,0)
\put(0,0){\epsffile{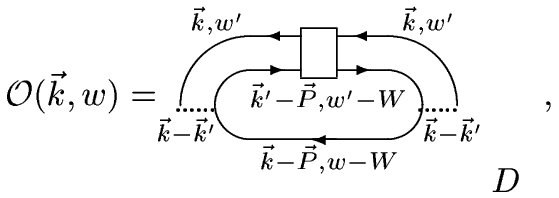}}\end{picture}
\end{eqnarray}
where the sub-diagram $\cal O$ is defined as an intermediate step.
This amounts to extend the MGE for the self energy of the quarks with
the simple one meson exchange.
Using the Weyl fermions, and expanding the ladder in meson poles, we
find that the self energy of the quark (antiquark) has a diagonal
component $\Sigma_d$ which contributes to the dynamical mass of the
quark (antiquark),
\begin{equation}
\label{4 diago}
\begin{picture}(127,25)(0,0)
\put(0,-10){\epsffile{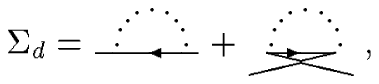}}\end{picture}
\end{equation}
and the energy of 1 quark is identical to the BCS one of
eq. (\ref{final 2 eqs}) except for the expected changes of the chiral
angle $\varphi$.
In eq.(\ref{4 diago}) we only included the nonvanishing diagrams
which remain from an expansion in powers of $1 / U$.
The free Green functions are proportional to $U^{-1}$.
The interactions without a pair creation or annihilation are proportional
to the infinite infrared constant $U$, while the remaining interactions
are finite.
It turns out that the new coupled channel diagrams vanish.
This happens because in the limit of an infinite $-U$ the box diagrams
in eq. (\ref{CS3 self}) which contribute to the quark
energy vanish. This is the case for instance of diagrams $(a)\, \ (c)$,
\FL
\begin{eqnarray}
\label{box}
\begin{picture}(220,40)(0,0)
\put(0,0){\epsffile{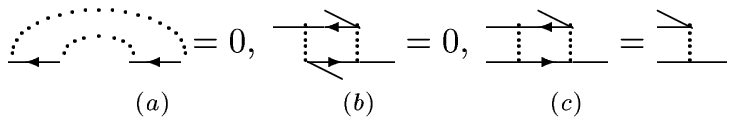}}\end{picture} .
\end{eqnarray}
We find that the quark energy $E=E_0-\Sigma_d$ remains the BCS one of
eqs. (\ref{2 eqs}) to (\ref{final 2 eqs}),
\FL
\begin{eqnarray}
E(k)&= &u^\dagger(k) \vec \alpha \cdot \vec k u(k)
- {1 \over 2 }\Bigl\{ u^\dagger (k)\int {d^3k' \over (2\pi)^3} V(k-k')
\nonumber \\ &&
\Omega_l\Bigl[ \Lambda^+(k') - \Lambda^-(k') \Bigr] \Omega_lu (k) \Bigr\}
\ .
\end{eqnarray}
However the last diagram of eq. (\ref{box}),
which contributes to the mass gap equation is finite.
\par
The mass gap equation is obtained when we impose that the antidiagonal
components $\Sigma_a$ of the self energy must cancel.
As in eq.(\ref{2 eqs}) this component in obtained with the projection
of the spinors $u^\dagger$ and $v$. This produces a function $\Sigma_a$
with the quantum numbers of a scalar, see eqs. (\ref{final 2 mgeqs})
 and (\ref{scalar wf gen}). In order to use
the results of the preceding section, it is convenient to fold
$\Sigma_a$ with a generic scalar wave function $\phi^+_{f_0}$. Then the
resulting product must vanish for any $\phi^+_{f_0}$. In fact this
ensures vacuum stability since this prevents the vacuum to decay
in scalar modes.
The diagrams that contribute to the antidiagonal component of
the self energy in the mass gap equation are now,
\FL
\begin{eqnarray}
\label{fold antidiago}
tr\left\{ {\phi^+_{f_0}(k)}^\dagger \Sigma_a \right\}=
\begin{picture}(120,25)(0,0)
\put(0,0){\epsffile{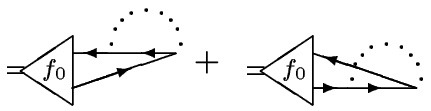}}\end{picture}
\nonumber \\
\begin{picture}(160,60)(0,0)
\put(0,0){\epsffile{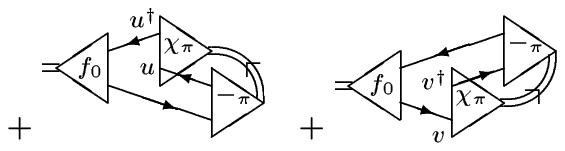}}\end{picture}
\nonumber \\
\begin{picture}(160,60)(0,0)
\put(0,0){\epsffile{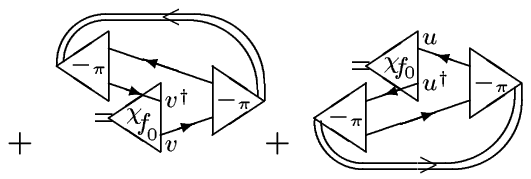}}\end{picture}.
\end{eqnarray}
In eq.(\ref{fold antidiago}) we only show
the diagrams which are nonvanishing in orders of $1 / U$, and in fact they
all are finite, of order $U^0$.
The first pair of diagrams are BCS diagrams.
It turns out that the new diagrams are the same diagrams which
contribute to the $f_0 \rightarrow \pi \otimes \pi$ coupling,
except for the negative energy wave function of the $\pi$ and
for the integral in the $\pi$ momentum $P$.
The negative energy wave-function $\phi^-$ always vanishes for
high momentum $P$ and in the case of low momentum $P$ it
is relevant only and for the pseudoscalar family of the $\pi$.
We suppose that the large number of excited states is not sufficient
to compensate the smallness of $\phi^-$, and we will not consider
this ultraviolet problem.
Thus we will only include the coupled channel contribution of the
$\pi$ family.
In the case of low momentum, the $\pi$ family has a extremely
large $\phi^-$, however the coupling to a quark $u^\dagger \chi u$
is derivative and vanishes. This prompts us to neglect the last line
of eq.(\ref{fold antidiago}). Finally we can remove, with a functional
derivative, the variational scalar wave function $\phi^+_{f_0}$.
The result after integrating in all the loop energies can be represented,
\FL
\begin{eqnarray}
\label{just momenta}
&&\begin{picture}(220,135)(0,0)
\put(0,0){\epsffile{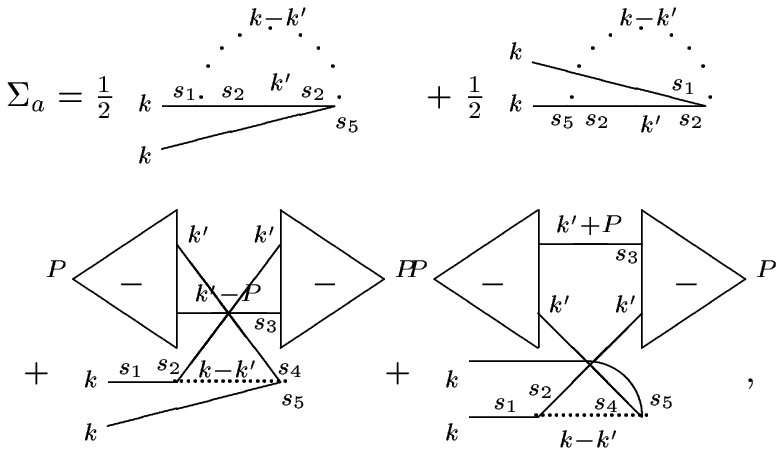}}\end{picture}
\end{eqnarray}
where the lines
\begin{picture}(20,0)(0,0) \put(0,0){\line(1,0){20}} \end{picture}
 only represent the spinors $u, \ v, \ u^{\dagger}$ or
$v^{\dagger}$, the integrals and the traces and no longer include
the quark or anti-quark propagators.
The mass gap equation $0={S^{-1}_0}_a-\Sigma_a$ is now,
\FL
\begin{eqnarray}
\label{mgapfin}
0&=& + u_{s_1}^\dagger(k) \vec \alpha \cdot \vec k v_{s_5}(k)
\nonumber \\ &&
-\Bigl\{ u_{s_1}^\dagger(k)
\int {d^3k' \over (2\pi)^3}  V (k-k')\Omega_l\Bigl[
\Bigr. \Bigr.
\nonumber \\ && \ \
+ u_{s_2}(k') \Bigl( { \delta_{s_2s_4} \over 2 } -
\int {d^3P \over (2\pi)^3} \phi_{s_2s_3}^-(P,k'-P/2) \Bigr.
\nonumber \\ && \ \ \ \ \ \
\Bigl. {\phi_{s_3s_4}^-}^{\dagger}(P,k'-P/2) \ \Bigr)
u_{s_4}^\dagger(k')
\nonumber \\ && \ \
- v_{s_2}(k') \Bigl({ \delta_{s_2s_4} \over 2 } -
\int {d^3P \over (2\pi)^3} {\phi_{s_2s_3}^-}^{\dagger}(P,k'-P/2) \Bigr.
\nonumber \\ && \ \ \ \ \ \
\Bigl. \phi_{s_3s_4}^- (P,k'-P/2) \ \Bigr)
v_{s_4}^\dagger(k')  \Bigl. \Bigr] \Omega_l v_{s_5}(k)  \, ,
\end{eqnarray}
where the sum over repeated spin indexes $s_i$ is assumed.
\par
%
%sssssssssssssssssssssssssssssssssssssssssssssssssssssss
%
\subsection{Model independent effects of the coupled channels}
\par
The dominant effect of
coupled channels is to multiply the potential term in the
mass gap equation by a factor of ,
\begin{equation}
\label{coupled channel factor}
1 - 2 \int{d^3P \over (2 \pi)^3} \phi^-_p {\phi^-_p}^{\dagger}  \, .
\end{equation}
This clearly decreases the term which is the source for the
spontaneous breaking of chiral symmetry. Thus the coupled
channel effect is to restore partially the chiral symmetry.
This effect is independent of the quark-quark interaction.
\par
The signs of the new terms in the mass gap equation deserve
a special attention since they determine whether the coupled
channel effect will increase or decrease the chiral condensation.
Because the coupled channel terms introduce in the mass gap equation a
new fermion loop it is natural for Dirac Fermions that
the coupled channel terms should be affected with a
$ - $ sign.
\par
When the Dirac fermions are translated into Weyl fermions
the quarks divide into the species of quarks and antiquarks which
have independent field operators and propagators, and the $ - $
signs are transfered from the propagator and the loops into the
antiquark vertex and the exchange diagrams,
and loops with quarks (antiquarks only).
In this case we check with Weyl fermions that the $-$ sign persists
and is due to the quark (antiquark) exchange. Only retardation, which
was not included here, might perhaps oppose to this negative sign.
\par
This sign can also be understood from the perspective of the
Mexican hat potential $- \lambda \sigma^2 +\mu \sigma^4$ of
effective meson models. In this case the quadratic term spontaneously
creates a scalar condensate, while the quartic term opposes to the
condensation and the actual condensate corresponds to the minimum
of the energy density where the two terms are balanced.
In the present paper there are three terms, a kinetic term which
opposes to the condensate (and has no correspondence in the effective
meson models) a BCS term which spontaneously breaks chiral
symmetry (it is equivalent to the quadratic term of effective meson
potentials) and a beyond BCS term which is equivalent to the quartic
term in effective meson potentials. This correspondence, which is
supported by the mean field theory where 
$\langle \bar \psi \psi \rangle \simeq \sigma$, 
confirms the negative sign of the coupled channel term.
Thus we may assume quite generally that coupled channels oppose
to the breaking of chiral symmetry.
\par
An interesting feedback from chiral symmetry to
the narrow width the groundstate occurs.
Chiral symmetry breaking can be understood variationally,
the solution $\varphi(k)$ of the mass gap equation also
minimizes the energy density $\cal E$ of the vacuum.
$\cal E$ is the sum of three terms, the free one, the
BCS one and the beyond BCS one.
Only the BCS term is negative and drives $\varphi(k)$
away from the trivial vanishing solution. The actual
solution $\varphi(k)$ minimizes the free term {\em and
minimizes the beyond BCS term} and at the same token produces
the most negative BCS term.
In eq.(\ref{fold antidiago}), we saw that the beyond BCS
diagrams in the mass gap equation are similar to the
diagrams of coupling of a scalar to a pair of pseudoscalars
(with low to moderate momentum).
Thus we conclude that this coupling is naturally suppressed
and that this suppression is selective in the sense that
it is not supposed to occur in other hadronic couplings.
This has an amplified effect in the scalar width which
is a function of the square of this coupling.
%
%ssssssssssssssssssssssssssssssssssssssssssssssssssssssssssssssssssssssss
%
\subsection{Solution of the mass gap equation}
\begin{figure}
\begin{picture}(403,150)(0,0)
\put(-53,-26){\epsffile{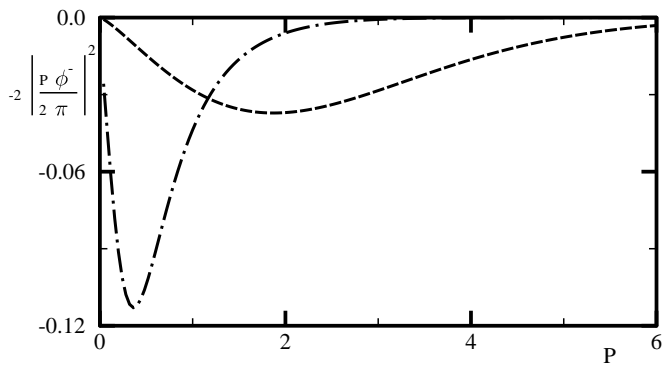}}\end{picture}
\caption{
We show the integrand of $\xi(0)$,
in the adimensional units of $K_0=1$.
The dotted line and dashed line correspond respectively to the
cases where $\phi$ is obtained in the low $P$ limit and in the high
$P$ limit.
\label{match}}
\end{figure}

We will now focus on the dominant terms among the coupled
channel contributions,
We obtain the $ momentum \otimes spin$ coupled channel contribution,
\begin{equation}
\label{xi spinmom}
\xi=\int d P dw
{-2 P^2 \phi^-(k+P/2)\phi^-(k+P/2)^\dagger  \over (2 \pi)^2}
\end{equation}
From the eqs. (\ref{pip0},\ref{large P wf}) we get for the
the integrand of eq. (\ref{xi spinmom}),
\FL
\begin{eqnarray}
\label{2 limits}
P \rightarrow 0 \ , \ \
&-0.14 \, { P^2 \over M(P)}
\left[S(k_1)-M(P){f_1(k_1) \over k_1} \right]^2
&\ ,
\nonumber \\
P \rightarrow \infty \ , \ \
&-1.9 \,
{a^2 P^{5/4} \over (1-a)^{11/4}} e^{- k_1^2/\alpha_p^2}
&\ ,
\end{eqnarray}
where the spin factor $ {i\sigma_2 \over \sqrt{2}}
{-i\sigma_2 \over \sqrt{2}}={1 \over 2} $ is included.
The integral that leads to $\xi$ is now evaluated with an
interpolation between the two limits of eq(\ref{2 limits}),
see Fig. \ref{match}.
Because this interpolation is arbitrary, we have to include in the
result a theoretical error.
%ffffffffffffffffffffffffffffffffffffffffffffffffffffffffffffffff
%fffffffffffffffffffffffffffffffffffffffffffffffffffffffffffffffff
\par
We obtain the $ momentum \otimes spin$ coupled channel contribution,
\begin{equation}
\xi (k) =  -(0.4 \pm 0.1 ) S
\end{equation}
which turns out to have a shape very close to the function
$S=sin(\phi)$ which was evaluated at the BCS level.
We estimate that the coupling to the quasi Goldstone bosons,
including the momentum, spin, color and flavor contributions
yields,
\begin{eqnarray}
\label{intphiphi}
\xi(k)
&\simeq & \ - \left( 0.4 \pm 0.1 \right)S 
{1 \over 3} \left( N_f -{1\over N_f}\right)
\nonumber \\
&\simeq& -( 0.3 \pm 0.2 ) S
 \ , \ \ N_f=2 \rightarrow 3  \ .
\end{eqnarray}
$N_f$ is the number of almost massless quark flavors which empirically
is between 2 and 3.
\par
The mass gap equation can be solved 
for a  coupled channels contribution equal to $\xi$, when $ \xi \geq -1$,
\FL
\begin{eqnarray}
0&=& u^\dagger \left[ \vec \alpha \cdot \vec k -{1 \over 2}
\Omega_l \int V(\Lambda^+ -\Lambda^-)(1+\xi)
\Omega_l \right] v
\\
&=&\Biggl\{
kS -\left[1-a(2C^2+1)\right]\left[{\Delta \varphi \over 2}
(1+\xi) + \dot \xi \dot \varphi\right] \Biggr.
\nonumber \\
&&\Biggl.
-SC \left[ {1\over k^2} +a \left( {\dot \varphi}^2- {1\over k^2}
\right) \right] (1+\xi)
+a SC \Delta \xi   \Biggr\}
\vec \sigma \cdot \hat k i \sigma_2
\nonumber
\end{eqnarray}
The solution $\varphi$ for $\xi =-0.3 \, S_{BCS} $ is shown
in Fig. \ref{chiral}.
In this case and for the same parameter $K_0$, quark condensate
$\langle \bar \psi \, \psi \rangle$ is decreased by a factor of  
$(1-0.12)^3$.
%
%ssssssssssssssssssssssssssssssssssssssssssssssssssssssssssssssssssssssssss
\section{Results and Conclusion}
We developed a general formalism to include both the effects
of chiral symmetry breaking and strong hadron-hadron interactions
in quark models.
This is encouraging since both effects are firmly established in
phenomenology.
We find new general effects in the scalar meson width, in the
breaking of chiral symmetry and in the mass shifts of the
hadron spectrum.
Quantitative results are computed within a model
which belongs to a class of Nambu and Jona Lasinio absolutely
confining instantaneous interactions, in the case where the coupled
pair of mesons are accounted as bare mesons.
\par
We find in this model that the mass and width of the light $q \bar q$
scalar $f_0$ meson are close \cite{papfpi} to the experimental mass
and width of the $f_0(980)$, and not to the $f_0(400-1200)$ or the
$f_0(1370)$.
This apparently indicates a possible solution to the scalar meson puzzle
without meson molecules, glueballs or strongly nonlinear coupled channel
effects.
In this case the attraction which is visible in $\pi \ \pi$
phase shifts and in the intermediate range $N N$ interactions would
need other interpretations \cite{Goldman} than the very wide
$\sigma$ meson.
\par
Compared with $\chi S B$ at the BCS level, a new parameter has been
identified, which leads to the percentage of coupled channel effects
in the mass gap equation. We find that coupled channels
suppresses the breaking of chiral symmetry.
This results are model independent. With our model we get
a suppression of the quark condensate by $5 \% \rightarrow 55 \%$
when the coupled channel effects are included.
\par
We find a new interesting feedback mechanism from chiral symmetry
to coupled channels and explain it variationally.
The chiral symmetry restorating contribution to the mass gap
equation from coupled channels is closely related 
to the coupling of a scalar to a pair of pseudoscalars.
The feedback enforces that the width of a groundstate scalar
decaying to a pseudoscalar pair (with low or moderate momentum)
is reduced when compared to the width of any other resonance.
This effect is model independent and contributes to understand
the scalar meson puzzle.
\par
Concerning real mass shifts we estimate that they are canceled
due to the new terms which are introduced by the Ward Identities.
This might improve previous \cite{Dullemond,ourcouca}
coupled channel calculations where this cancellation was not
explicitly included.
An systematic shift of the hadron spectrum is not expected.
In this sense we agree with the results of Geiger and Isgur
\cite{Isgur}.
Nevertheless mass splittings between states with different quantum
numbers are still expected from the boundstate equation.
Different theoretical problems that could be reviewed with these
new techniques are the contribution of the coupled channels to the
hadron spectra, for instance to the interesting $\eta'$ mass or to
the $N$ mass.
\par
We conclude that in general the results of this paper, without
ruling out other perspectives, explain why the simpler quark
model is so successful.
\acknowledgements
I am very grateful to Emilio Ribeiro for long discussions and
suggestions since 1989 on the pion mass problem in relation with
the coupled channels, on the BCS
mechanism and on Ward identities.
I also acknowledge George Rupp, Jean-Marc Richard, Jack Paton,
Robin Stinchcombe, Nils Tornqvist, Vitor Vieira, Pedro Sacramento
and Adam Szczepaniak for comments or suggestions.
%
%
%aaaaaaaaaaaaaaaaaaaaaaaaaaaaaaaaaaaaaaaaaaaaaaaaaaaaaaaaaaaaaaaaaa
\appendix
%
%aaaaaaaaaaaaaaaaaaaaaaaaaaaaaaaaaaaaaaaaaaaaaaaaaaaaaaaaaaaaaaaaaaaaaa
\section{Confinement with infrared finite coupled channels}
At the BCS level, the boundstates are obtained with the ladder
approximation which is equivalent to the Salpeter equation,
and to the Schr\"odinger equation (see Appendix C).
In this case the infinite infrared divergent constant $U$
of the interaction (\ref{potential},\ref{interaction})
is extremely convenient \cite{Gastao} to remove the colored states, which
have masses proportional to $U$.
Let us consider the dominant terms in orders of $U$
of the energy of a system of $n$ quarks and antiquarks.
The 1-body energy includes the self energy
(\ref{final 2 eqs}) which is calculated
with the Schwinger Dyson equation,
\begin{eqnarray}
\label{U cancel}
\sum_i E_i +\sum_{i<j} V_{ij} &\simeq&
{3 \over 4} \left( { U \over 2} \sum_i
{\vec \lambda _i \over 2} \cdot {\vec \lambda _i \over 2}
+U\sum_{i<j}{\vec \lambda _i \over 2} \cdot {\vec \lambda _j \over 2}
\right)
\nonumber \\ &\simeq&
{ 3 \over 32 } U \vec \Lambda \cdot \vec \Lambda \ ,
\end{eqnarray}
where $\vec \Lambda$ is the Gell Mann matrix of the total color of the
system.
The energy (\ref{U cancel}) vanishes for color singlets only,
thus colored states have an infinite energy and are confined.
\par
To include coupled channels in the energy of a color singlet,
for instance in a meson, we consider the
complete series of diagrams that contribute to the irreducible
$q \, \bar q$ interaction.
One has to include all the possible number of quark
loops and all the possible insertions of the microscopic quark-quark
interaction.
Then this series can be resumed in order to factorize the bare meson
($q \, \bar q$)
and hadron ($q \, q \, q$) ladders.
According to Appendix C, the ladder needs
integrals $\int {dw \over 2 \pi}$ in all the external relative energies
in order to have a hadron pole.
To ensure in a particular diagram that the
meson poles are present it is convenient to decompose the ladder in,
\FL
\begin{equation}
\label{ladder resum}
\begin{picture}(170,27)(0,0)
\put(0,0){\epsffile{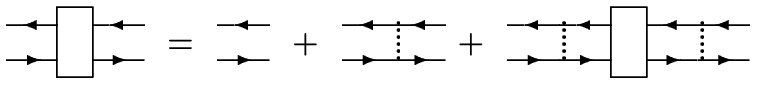}}\end{picture}
\end{equation}
where in the right hand side the ladder is integrated and contains
the meson pole, and the first pair of diagrams contributes
to the overlap interactions of hadrons.
In order to simplify the calculations it is convenient to truncate
eventually the series of diagrams.
The pertubative parameter is then the number of considered ladders.
This is both straightforward and also close to the hadronic
phenomenology.
\par
We now show how the interactions of color singlet ladders
can be finite, when they are built from the infrared divergent
quark-quark microscopic interaction (\ref{interaction}), and
the quark propagators are divergent as well.
When a hadron is emitted or absorbed we have
quark-antiquark annihilation overlaps, for instance
in a 3-meson vertex,
\FL
\begin{equation}
\label{annihilation overlaps}
\begin{picture}(170,50)(0,0)
\put(0,0){\epsffile{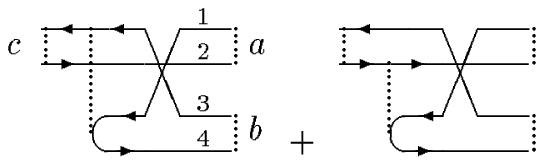}}\end{picture}
\end{equation}
We first integrate the relative energies in the first diagrams of q.
(\ref{annihilation overlaps}),
\FL
\begin{eqnarray}
&&\int {dw dw' \over (2\pi)^2} {i \over w -E_1 + i \epsilon} iV_{1\bar 4,3}
{i \over -w' -E'_4 + i \epsilon} \nonumber \\ &&
\ ={i \over w' -E'_3 + i \epsilon}
{i \over w -E_3 + i \epsilon}
{i \over -w -E_2 + i \epsilon} \nonumber \\ &&
\ = -iV_{1\bar 4,3} {G_0}_a{G_0}_b{G_0}_c\ .
\end{eqnarray}
where the ${G_0}_n$ are in fact part of the respective BS amplitudes
which are untruncated.
The remaining factor $-iV_{1\bar 4,3}$ is finite because the
term proportional to  $U$ in the quark-antiquark annihilation vertex
is $v^\dagger_{s}(k)\delta^3(k-k')u_{s'}(k')=0$.
\par
When the number of hadrons are conserved, this is
the case in elastic scattering, we have quark exchange overlaps,
for instance in a 4-meson vertex,
\FL
\begin{eqnarray}
\label{exchange overlaps}
\begin{picture}(220,50)(0,0)
\put(0,0){\epsffile{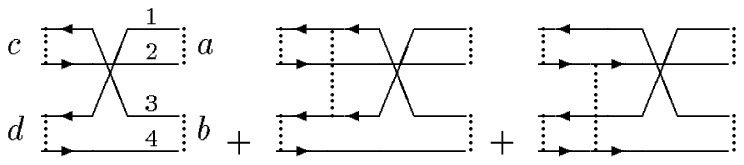}}\end{picture}
\end{eqnarray}
The integrals in the relative energies are,
\FL
\begin{eqnarray}
&&\int {dw \over 2\pi} {i \over w -E_1 + i \epsilon}
{i \over -w -E_4 + i \epsilon} {i \over w -E_3 + i \epsilon}
{i \over -w -E_2 + i \epsilon}
\hspace{-.2cm} \nonumber \\ && \
= i(E_1+E_2+E_3+E_4){G_0}_a{G_0}_b{G_0}_c{G_0}_d    \nonumber \\ && \
\int {dw dw' \over (2\pi)^2} {i \over w -E_1 + i \epsilon} (-iV_{13})
{i \over w' -E'_1 + i \epsilon}  \nonumber \\ && \
{i \over -w' -E'_4 + i \epsilon}
{i \over w' -E'_3 + i \epsilon}
{i \over w -E_3 + i \epsilon}
{i \over -w -E_2 + i \epsilon} \nonumber \\ && 
\ = -iV_{13} {G_0}_a{G_0}_b{G_0}_c{G_0}_d \ .
\end{eqnarray}
The four ${G_0}_n$, will be absorbed by the Salpeter amplitudes
and the remaining factors are,
\begin{eqnarray}
&i(E_1+E_2+E_3&+E_4-V_{13}-V_{24})P_{13}  \nonumber \\
&&=i(E_1+E_2+E_3+E_4+V_{14}+V_{23})P_{13} \nonumber \\
&&=iP_{13}(E_1+E_2+V_{12}+E_3+E_4+V_{34}) \nonumber \\
&&=iP_{13}(0U+\dots)
\end{eqnarray}
where the infinite $-U$ terms cancel in the same way as
in eq. (\ref{U cancel}) because the mesons
$_{32}$ and $_{14}$ in the left and $_{12}$ and $_{34}$
in the right are color singlets.
We find that all the terms proportional to $U$ cancel when the
complete set of diagrams which contribute
to the interaction between color singlets is included.
\par
In this framework the masses of bare hadrons and their
interactions are finite and can be evaluated.
Then we compute the masses and widths of dressed hadrons.
These are the final freedom degrees which can be compared
with the experimental spectrum of hadronic resonances.
%
%aaaaaaaaaaaaaaaaaaaaaaaaaaaaaaaaaaaaaaaaaaaaaaaaaaaaaaaaaaaaaaaaaaaaaaa
\section{The light $\pi$ and Ward Identities}
The solution to the pion mass problem is found using
\cite{Adler,Schrieffer} the Ward identities (WI) in order to
insure that the bound state equation for the pion
-a Bethe Salpeter equation with coupled channels -
is consistent with the non linear mass gap equation.
The WI were first derived for fermion-gauge field theories, and were
initially based on the simple observation that for free fermions with
propagator $\displaystyle {\cal S}_0(p)= {i \over \not p-m_0}$
and a free vector vertex $\Gamma^{\mu}_0= \gamma^{\mu}$,
\FL
\begin{equation}
\label{vector WI}
i (p_{\mu}-p_{\mu}') {\cal S}(p)\Gamma^{\mu}(p,p'){\cal S}(p')
= {\cal S}(p) -{\cal S}(p')   
\end{equation}
The difference in the right member of the equation
extends the identity to renormalized propagators
and vertices.
This identity is then crucial for the conservation
of electric charge.
The WI enforce that
the self energy of the MGE is obtained (without double-counting)
from the BS kernel by closing the fermion line where
the vertex is inserted.
Inversely, they also ensure
that the BS kernel is obtained if one inserts the vertex in all possible
propagators of the self energy.
For instance this mapping is trivial at the BCS level
where the mass gap equation (\ref{BCS MGE}) is clearly equivalent to
the bound  state equation (\ref{BCS BSE}).
Let us now consider a more general case, where the fermion self energy
include a product of bare propagators $S_{\alpha_i \beta_i}(k_i +P)$.
The external momentum is $P$ and $k_i$ is a loop momentum.
The propagators can be factorized and we get,
\begin{equation}
\Sigma(P) = \cdots \prod_i S_{\alpha_i \beta_i}(k_i +P)  \ .
\end{equation}
Then the vertex $\Gamma$ can be constructed if we insert a bare vertex
in all possible bare propagators, and we get,
\begin{eqnarray}
\Gamma^\mu(P_1-P_2) &=& \cdots \sum_j \prod_{{i1} \leq j} S_{\alpha_{i1} \beta_{i1}}(k_{i1}
+P_1)  \nonumber \\ && \gamma^\mu \ \
\prod_{{i2} \geq j} S_{\alpha_{i2} \beta_{i2}}(k_{i2} +P_2) \ .
\end{eqnarray}
Inversely the original self energy can be recovered if we substitute the
bare vertex by the difference of propagators,
\begin{eqnarray}
\label{vertex self}
i({P_1}_\mu-{P_2}_\mu) \Gamma^\mu
 &&= \cdots \sum_j \prod_{{i1} < j} S_{\alpha_{i1} \beta_{i1}}(k_{i1} +P_1)
\nonumber \\
\left[ S_{\alpha_j \beta_j}(k_j +P_1)- \right.&&
\left. S_{\alpha_j \beta_j}(k_j +P_2)\right]
\prod_{{i2} > j} S_{\alpha_{i2} \beta_{i2}}(k_{i2} +P_2)
\nonumber \\
&&= \left[ \Sigma(P_1) - \Sigma(P_2) \right ]
\end{eqnarray}
where the products which depend on both $P_1$ and $P_2$ cancel.
and this removes the double counting.
\par
There is also a WI identity for the free axial vector vertex
$\Gamma^{\mu 5}_f= \gamma^{\mu}\gamma^5$ that involves the free
pseudoscalar vertex $\Gamma^5_f= \gamma^5$,
\FL
\begin{eqnarray}
\label{axial WI}
&&-i (p_{\mu}-p_{\mu}') {\cal S}(p)\Gamma^{\mu 5}(p,p'){\cal S}(p')
\nonumber \\ &&+
2im{\cal S}(p)\Gamma^5(p,p'){\cal S}(p')
= {\cal S}(p)\gamma^5 + \gamma^5{\cal S}(p')   
\end{eqnarray}
which is valid in a renormalization program providing the interaction
is chiral invariant. In this case an equation analogous with
eq.(\ref{axial WI}) is complied by the axial and pseudoscalar
vertices and by the self energy.
\par
A key product of the axial WI is the proof \cite{Nambu,Pagels}
that a pseudoscalar Goldstone boson exists when current quark
are vanishing and chiral symmetry breaking occurs.
The full propagator is then renormalized and the self-energy
$\Sigma$ has a mass-like term,
\begin{equation}
{\cal S}^{-1}(p)={\cal S}_0^{-1}(p)-\Sigma(p) \ , \ i \Sigma =
 A(p)-\not p B(p)  \ .
\end{equation}
\begin{equation}
\label{self}
{\cal S}^{-1}(p)={\cal S}_0^{-1}(p)-\Sigma(p) =
{A(p)\not p - B(p) \over i} \ .
\end{equation}
If we substitute this propagator in the WI, we find the solution for
the pseudoscalar vertex $\Gamma^5$ with a vanishing $p-p'$,
\begin{equation}
\label{amplitude}
\Gamma^5(p = p') = { B(p) \over m_0 } \gamma^5
\end{equation}
which diverges for a vanishing quark mass $m_0$ and shows
that the pole of a massless pseudoscalar meson appears in the
axial vector vertex, with a boundstate truncated amplitude of
\begin{equation}
\chi(p-p'\simeq 0) ={B \gamma_5 \over f_\pi}
\end{equation}
where $f_\pi$ includes a norm.
Incidentally, this identity also offers a proof of the Gell-Mann,
Oakes and Renner relation.
If we expand the vertex $\Gamma^5$ in the neighborhood of the $\pi$
pole,
\FL
\begin{eqnarray}
\Gamma^5=\chi {i \over (p-p')^2 -M^2}
\, tr \hspace{-.03cm}\left\{\chi {\cal S}(p)\Gamma^5_0 {\cal S}(p')\right\}
\end{eqnarray}
where we included the integral in the trace.
substituting $\Gamma_5$ in eq. (\ref{amplitude}) and performing a trace
with ${\cal S}\gamma_5{\cal S}$, we find for small masses
\begin{eqnarray}
&& \, tr \hspace{-.03cm} \left\{ \gamma_5 {B {\cal S}\gamma_5 {\cal S}\over f_{\pi} } \right\}
{i \over -M^2}
\, tr \hspace{-.03cm} \left\{ {B{\cal S}\gamma_5 {\cal S}\over f_{\pi} } \gamma_5 \right\}
= \, tr \hspace{-.03cm} \left\{\gamma_5 {B {\cal S}\gamma_5 {\cal S}\over m_0} \right\}
\nonumber \\
&&\Rightarrow
-m_0 \, tr \hspace{-.03cm}\left\{{\cal S}\right\}
=-2m_0 \langle \bar \psi \, \psi \rangle = M^2 f_{\pi}^2
\end{eqnarray}
\par
In QCD it is necessary to include the axial anomaly, in the flavor
singlet WI which corresponds to the $\eta'$ channel.
The flavored axial currents remain unchanged,
in particular the $\pi, \, K,  \, \eta$ remain quasi Goldstone bosons.
We will now simply assume that the $\eta '$ is heavier
\cite{Swanson,Nambu,Tornqvist,Hooft} than
the usual $N_f^2-1$ Goldstone bosons, where $N_f$ is the number of light
quark flavors.
%
%aaaaaaaaaaaaaaaaaaaaaaaaaaaaaaaaaaaaaaaaaaaaaaaaaaaaaaaaaaaaaaaaaaaaaaa
%
\section{The Salpeter equations in the Energy-Spin formalism}
When the interaction is instantaneous, a simplification occurs
in the Bethe-Salpeter S matrix in the ladder approximation,
\FL
\begin{eqnarray}
\label{4 S matrices}
\begin{picture}(220,180)(0,0)
\put(0,0){\epsffile{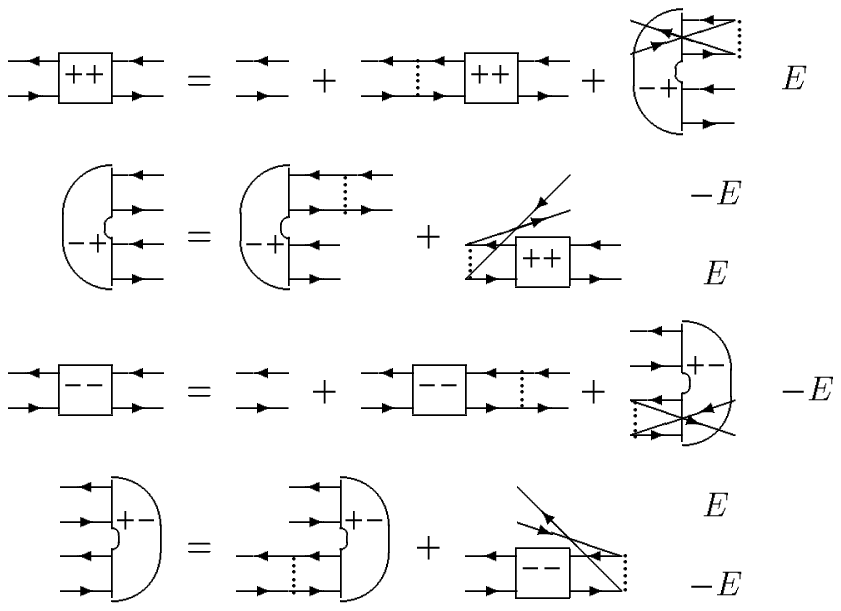}}\end{picture} .
\end{eqnarray}
The S matrix only has $4$ independent sub-matrices, which have to be 
calculated iteratively.
The other $12$ sub-matrices are directly computed from the independent 
$4$ ones. The Salpeter equations are obtained when all the relative
quark-antiquark energies
of the system (\ref{4 S matrices}) are integrated,
\begin{eqnarray}
&&\int {dw \over 2 \pi} {i \over w + W/2 -E_q + i \epsilon}
{i \over w + W/2 -E_{\bar q} + i \epsilon}=
\nonumber \\
&&{i \over W -E_q -E_{\bar q}+ i \epsilon} \ ,
\end{eqnarray}
The Salpeter
equations can then be written in the compact form,
\FL
\begin{eqnarray}
\label{compact}
S&=&G_0+G_0VS
\nonumber \\ &\Rightarrow & (G_0^{-1}-V)S = 1
\nonumber \\ &\Rightarrow & (W\sigma_3-H+ i \epsilon)S=i
\end{eqnarray}
where $ \sigma_3 $ is the Pauli matrix,
\begin{eqnarray}
S&=&\left[\begin{array}{cc}
S^{++}&S^{+-}\\
S^{-+}&S^{--}
\end{array}\right] \ ,
\nonumber \\
G_0&=&\left[\begin{array}{cc}
{i \over W - E_q -E_{\bar q} + i \epsilon} & 0 \\
0 & {i \over -W  - E_q -E_{\bar q} + i \epsilon}
\end{array}\right] \ ,
\nonumber \\
V&=&-i\left[\begin{array}{cc}
\int V_d&\int V_a\\
\int V_a&\int V_d
\end{array}\right] \ ,
\nonumber \\
H&=&\left[\begin{array}{cc}
E_q +E_{\bar q} +\int V_d&\int V_a\\
\int V_a&E_q +E_{\bar q} +\int V_d
\end{array}\right]
\end{eqnarray}
and it turns out that $H$ is an hermitian and positive operator.
\par
The Salpeter wave-functions are the solutions of the homogeneous equations,
\begin{equation}
\label{homogeneous}
(M\sigma_3-H)\phi = 0 \ ,
\phi =\left(\begin{array}{c}
\phi^+\\
\phi^-
\end{array}\right)
\end{equation}
which is an eigenvalue equation,
similar to the Schr\"o- dinger equation,
except for the $\sigma_3$ .
This formalism is known as
the energy-spin formalism, where $\phi^+$ is called the positive energy
wave-function and $\phi^-$ is called the negative energy wave-function.
The Salpeter equation (\ref{homogeneous})
is equivalent to the variational equation,
\begin{equation}
\delta \left({\langle \phi | H | \phi \rangle
\over \langle \phi | \sigma_3 | \phi \rangle} \right) =0 \ ,
\end{equation}
which suggests that a normalizing condition of the wave functions
might be,
\begin{equation}
\langle \phi | \sigma_3 | \phi \rangle =|\phi^+|^2 - |\phi^-|^2=1 \ .
\end{equation}
Let us study the class of solutions $\phi_u$ where this norm is possible,
i. e. $|\phi^+_u| > |\phi^-_u|$. Once the single quark energies $E$ and the
2-quark diagonal $V_d$ and antidiagonal $V_a$ potentials are defined,
the solutions can be obtained numerically, either iteratively or
variationally, and one finds a whole
spectrum of solutions with energy $M>0$. When $M$ increases, we find that
$|\phi^-_u|$ is proportional to $M^{-1}$, and in the limit of large mass
we exactly recover Schr\"odinger equation. However another class of solutions
$\phi_d$ is unavoidable, with a one to one correspondence with the $\phi_u$,
and where $|\phi^-_d| > |\phi^+_d|$,
\begin{eqnarray}
\phi^+_d=\phi^-_u \ , \ \phi^-_d=\phi^+_u \ , \ M_d = - M_u \nonumber \\
\langle \phi_u | \sigma_3 | \phi_u \rangle = 1 \ , \
\langle \phi_d | \sigma_3 | \phi_d \rangle = -1
\end{eqnarray}
thus the spectrum is unbound. Exactly half of the solutions have
a negative mass and a negative norm.
When we include the infinitesimal $i \epsilon$ in eq. (\ref{homogeneous}),
we find that the larger component of $\phi$ dominates and the eigenvalues
are now,
\begin{eqnarray}
M_u &\rightarrow& M_u -i \epsilon  \nonumber \\
M_d &\rightarrow& M_d +i \epsilon = - (M_u -i \epsilon)
\end{eqnarray}
\par
The operators $H$ and $\sigma_3$ are both hermitian, thus the
set of solutions $|\phi\rangle$ constitute a basis of the Hilbert
space, orthogonal in the sense that $ \langle \phi | \sigma_3 |
\phi '\rangle =0$, and the identity element is,
\begin{equation}
1=\sum { \sigma_3 | \phi \rangle    \langle \phi |
\over  \langle \phi |\sigma_3 | \phi \rangle }
=\sum_u  \sigma_3 | \phi \rangle    \langle \phi |
-\sum_d  \sigma_3 | \phi \rangle    \langle \phi |
\end{equation}
Inserting this partition in the $S$ matrix equation (\ref{compact}),
we find,
\FL
\begin{eqnarray}
i&=& (W\sigma_3-H)S \nonumber \\
&=& \sum { \sigma_3 | \phi \rangle    \langle \phi |
\over  \langle \phi |\sigma_3 | \phi \rangle }
(W\sigma_3-H+i \epsilon)\sigma_3
\sum { \sigma_3 | \phi \rangle    \langle \phi |
\over  \langle \phi |\sigma_3 | \phi \rangle }
\sigma_3 S \nonumber \\
&=& \sum \sigma_3 | \phi \rangle {W-M
+i \epsilon \langle \phi |\sigma_3 | \phi \rangle
\over \langle \phi |\sigma_3 | \phi \rangle }
\langle \phi |\sigma_3 S\nonumber \\
\sigma_3S &=& \sum \sigma_3 | \phi \rangle {i
\langle \phi |\sigma_3 | \phi \rangle  \over W-M
+i \epsilon \langle \phi |\sigma_3 | \phi \rangle
} \langle \phi | \nonumber \\
S &=& \sum_u | \phi_u \rangle {i \over W-M_u+i\epsilon } \langle \phi_u |
\nonumber \\
&&- \sum_d | \phi_d \rangle {i \over W-M_d-i\epsilon } \langle \phi_d |
\end{eqnarray}
and the states with negative energy and negative norm can be reinterpreted as
boundstates with positive mass moving forward in time, where the variable
$W=P_0$ in the propagator turns out to be negative,
\begin{eqnarray}
S &=& \sum_u | \phi_u \rangle {i \over W-M_u+i\epsilon } \langle \phi_u |
\nonumber \\
&&+ \sigma_1 | \phi_u \rangle {i \over -W-M_u+i\epsilon }
\langle \phi_u | \sigma_1
\end{eqnarray}
and it suffices to work with the $\phi_u$. Diagrammatically we get for
instance,
\FL
\begin{eqnarray}
\label{last ladder}
\begin{picture}(200,60)(0,0)
\put(0,0){\epsffile{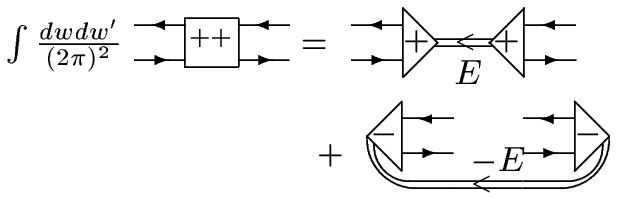}}\end{picture}
\end{eqnarray}
where $w$ and $w'$ are the external energies.
The quark antiquark BS amplitude is represented by the triangle
\begin{picture}(25,10)(0,0)
\put(0,8){\line(1,0){5}}
\put(0,2){\vector(1,0){10}}
\put(15,8){\vector(-1,0){10}}
\put(15,2){\line(-1,0){5}}
\put(15,0){\line(0,1){10}}
\put(15,0){\line(2,1){10}}
\put(15,10){\line(2,-1){10}}
\end{picture}
and the boson propagator of the meson is represented by the double line
\begin{picture}(15,2)(0,0)
\put(0,2){\line(1,0){15}}
\put(0,0){\line(1,0){15}}
\end{picture}.
It is consistent to substitute the $_d$ wave-functions and masses
in terms of the components $\phi_u^+ \ , \ \phi_u^-$ and the
masses $M_u$. In the paper this reinterpretation will be assumed and
we will skip the subindex $_u$.
The practical result is that we can use the amplitude $\phi^+$ when the
bra(ket) precedes(succeeds) the meson propagator, and we use the amplitude
$\phi^-$ for the opposite case.
%
%aaaaaaaaaaaaaaaaaaaaaaaaaaaaaaaaaaaaaaaaaaaaaaaaaaaaaaaaaaaaaaaaaaaa
\section{The Bethe Salpeter equation with coupled channels}
We now go beyond BCS including the mesonic coupled channels
both in the MGE and in the bound state equations.
It is convenient to return to the Dirac formalism in order to
reduce the number of diagrams when we apply the WI.
We will extend the BS for the $q \bar q$ boundstate of the quarks
with the minimal meson loop of coupled channels.
\FL
\begin{eqnarray}
\label{CS3 tentative}
\begin{picture}(220,60)(0,0) %248 in dia33.ps
\put(0,0){\epsffile{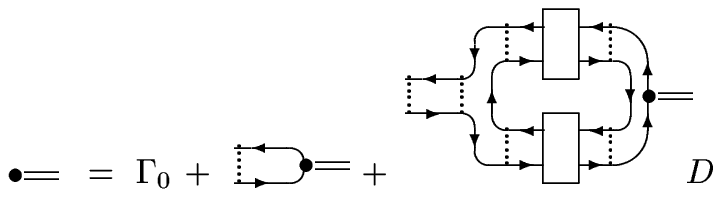}}\end{picture}
\end{eqnarray}
\par
where we follow the approximation of including only 3-legged
effective meson vertices in the meson interaction. We will
show that this is imposed by the box at the left of the beyond
BCS diagram in eq. (\ref{CS3 tentative}).
Upon iteration the equation,
\begin{equation}
\Gamma = \Gamma_0 + V \, G_0 \, \Gamma + V \, G_0 \, V \,
M \Gamma
\end{equation}
can be resumed.
If we factorize the S matrix at the ladder level
$S_0=(1-G_0V)^{-1}G_0$, see eq. (\ref{compact}),
the ladder will appear in the middle of the coupled channel terms,
\begin{eqnarray}
G_0 \,\Gamma
&=& S_0 \, \Gamma_0 + G_0  V \, S_0 \, V \, M \Gamma \ ,
\nonumber \\
\Gamma &\simeq& V \, S_0 \Gamma_0
+ V \, S_0 \, V \, M \Gamma.
\end{eqnarray}
When the ladders $S_0$, including the two ones of the meson loop in
$M$, are expanded in meson poles and wave functions according to
eq. (\ref{last ladder}), we recover the meson - meson pair
coupling of eq. (\ref{scalar coupling}).
The resulting pole of the vertex $\Gamma$ is the mass of the
dressed meson $M_0+\Delta M$.
This is equivalent to the Resonating Group Method equations 
\cite{Hecht} for coupled channels of one meson with a pair of 
non-interacting mesons.
\par
We now use the WI prescription of removing the vertex and closing the
respective fermion line to arrive at the mass gap equation
(\ref{CS3 self}). To recover the full vertex equation we must
insert the vertex in all possible propagators of eq. (\ref{CS3 self}).
We then arrive at a WI consistent Bethe Salpeter equation for the
vertex and for the boundstate,
\FL
\begin{eqnarray}
\label{CS3 kernel}
\begin{picture}(220,300)(0,0) %248 in dia32.ps
\put(0,0){\epsffile{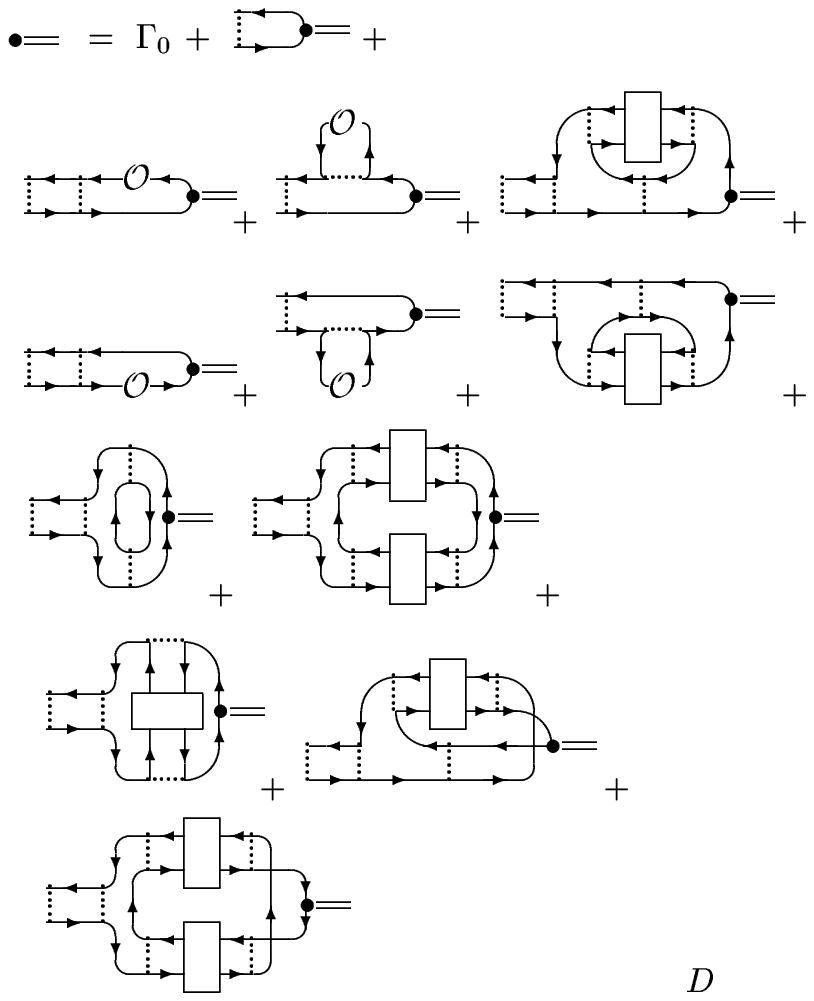}}\end{picture}
\end{eqnarray}
where the diagrams are shown in separate lines according to their
properties.
Line 1 corresponds to the BS equation at the BCS level.
Without the other lines it would reproduce the ladder Bethe-Salpeter
equation for the vertex.
The lines 2, 3 and 5
were separated because they contain all the contain terms
proportional to the infrared divergent $U$,
but they cancel in each line and all lines are finite.
The remaining lines 4 and 6 contain the terms that one
would except in coupled channel equations  where a pair
of mesons is created and then annihilated (except for the
first diagram of line 4 which vanishes). With them we
calculate for instance the partial decay width of a resonance
into a channel of 2 mesons.
The last lines 5 and 6 are only relevant for flavor singlets
because the quark pair in the incoming meson is annihilated,
thus for flavor vectors they are null.                    
\par
The cancellation of the infrared divergences become clear in the
Goldstone-Weyl formalism. Let us consider for instance the diagrams
of line 2 in eq. (\ref{CS3 kernel}),
\FL
\begin{eqnarray}
\begin{picture}(220,50)(0,0)
\put(0,0){\epsffile{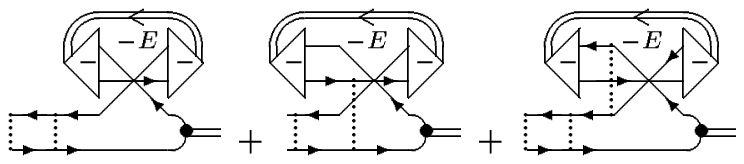}}\end{picture}
\end{eqnarray}
These 3 diagrams are infrared divergent but their sum is finite,
in an analogous way to eq. (\ref{exchange overlaps}).
\par
It is important to remark that in the previous calculations in
the literature, the extrapolation from the ladder level to the
coupled channel level would only include the diagrams of lines 4 and 6.
We now find the previous choice arbitrary since the diagrams of lines
2, 3 and 5 were not considered.
The role of these diagrams is to cancel any real mass shift of the
$\pi$ due to the usual coupled channel diagrams of lines 4 and 6,
in order that the $\pi$ remains a Goldstone boson in the chiral limit.
This is ensured by the WI, see eq. (\ref{amplitude}).
Concerning widths, since the new diagrams are real, the results
of the previous calculations in the literature are correct.
However we find no systematic real mass shift in the meson spectrum due
to coupled channels. This contradicts most of the real mass shifts of
hundreds of $MeV$ which are common in the literature.
Only splittings between different levels, due to neighboring cuts,
may be affected by coupled channels.
%
%rrrrrrrrrrrrrrrrrrrrrrrrrrrrrrrrrrrrrrrrrrrrrrrrrrrrrrrrrrrrrrrrrrrrrrrrrrrr

%

\begin{references}
%
\bibitem{PDB} Review of Particle Physics, Particle Data Group,
Phys. Rev. D {\bf 54},1 (1996).
%
\bibitem{Swanson} A. Szczepaniak, E. Swanson, C.-R. Ji and S. Cotanch,
Phys. Rev. Lett. {\bf 76}, 2011 (1996).
%
\bibitem{Weinstein}J. Weinstein, N. Isgur, Phys. Rev. Lett {\bf 48},
659 (1982); Phys. Rev. D {\bf 27}, 588 (1983).
%
\bibitem{Dullemond} E. Van Beveren, T. Rijken, K. Metzger, C. Dullemond,
G. Rupp, J. Ribeiro  Zeit. Phys. C {\bf 30}, 615 ,(1986).
%
\bibitem{Torsigma}N. T\"ornqvist, M. Roos, Phys. Rev. Lett {\bf 76}, 1575,
(1996); N. Tornqvist Zeit. Phys. C {\bf 68}, 647 (1995).
%
\bibitem{BCS}
J. Bardeen, L. N. Cooper and J. R. Schrieffer, Phys. Rev. {\bf 108},
1175 (1957); L. V. Keldysh and A. N.
Kozlov, Sov. Phys. JETP {\bf 27}, 521 (1968).
%
\bibitem{Nambu}
Y. Nambu, G. Jona-Lasinio, Phys. Rev. {\bf 122}, 345 (1961);
{\bf 124}, 246 (1961).
%
\bibitem{bBCS in CSB}
F.T. Hawes and A. G. Williams, Phys. Lett. B
{\bf 268}, 271 (1991); A. Bashir and M. R. Pennington, Phys. Rev.
D {\bf 50},
7679 (1994); H. J. Munczek, Phys. Rev. D {\bf 52}, 4736 (1995);
A. Bender, C. D. Roberts, L. V. Smekal, Phys. Lett. B {\bf 380}, 7 (1996).
%
\bibitem{preprint}
P. Bicudo, preprint FISIST/5-97/CFIF and hep-ph/9703229  (1997).
%
\bibitem{3P0}
P. Bicudo and J. Ribeiro, Phys.\ Rev.\ D {\bf 42}, 1611 (1990).
%
\bibitem{papfpi} Y. le Yaouanc, L. Oliver, S. Ono, O. Pene and
J.-C. Raynal, Phys. Rev. D {\bf 29}, 1233 (1984);
Phys. Rev. D {\bf 31}, 137 (1985).
%
\bibitem{ourcouca}
P. Bicudo and J. Ribeiro, Phys. Rev. D{\bf 42}, 1635 (1990);
P. Bicudo, J. Ribeiro and J. Rodrigues, Phys. Rev. C{\bf 52},
2144 (1995);
P. Bicudo and J. Ribeiro, Phys. Rev. C {\bf 55}, 834, (1997).
P. Bicudo, L. Ferreira, C. Placido and J. Ribeiro,
Phys. Rev. C {\bf 56}, 670, (1997).
%
\bibitem{newrho}
R. Friedrich and H. Reinhardt, Nucl. Phys. A{\bf 549}, 406 (1995);
S. Gao, C. Shakin and W.-D. Sun, Phys. Rev. C {\bf 53}, 1374 (1996);
K. Mitchel and P. Tandy, Phys. Rev. C {\bf 55},1477 (1997);
C. Shakin and W.-D. Sun, Phys. Rev. D {\bf 55}, 2874 (1997).
%
\bibitem{Richard}
S. Pepin, F. Stancu, M. Genovese, J. M. Richard,  to be published in
COMO2, world scientific, preprint nucl-th/9608058.
%
\bibitem{Thomas} A. Thomas and G. Miller, Phys. Rev. D {\bf 42}, 288
(1991).
%
\bibitem{Tornqvist}N. T\"ornqvist, preprint hep-ph/9612238 (1996);
Phys. Lett. B{\bf 406}, 70 (1997);
Phys. Lett B{\bf 426}, 115 (1998).
%
\bibitem{charm}
J. E. Villate, D.-S. Liu, J. E. Ribeiro, P. Bicudo, Rev. D
{\bf 47}, 1145 (1993).
%
\bibitem{Roberts}
R. Cahill and C. Roberts, Phys. Rev. D {\bf 32}, 2419 (1985).
%
\bibitem{Liu}
Y. Dai, Z. Huang and D. Liu, Phys. Rev. D {\bf 43}, 1717 (1991).
%
\bibitem{Adler} S. Adler and A. C. Davis, Nucl. Phys.
B {\bf 224}, 469 (1984).
%
\bibitem{Kalinowsky}
Y. Kalinowsky, L. Kaslun, and V. Pervushin, Phys. Lett. B{\bf 231},
288 (1989);
R. Horvat, D. Kekez, D Klabucar and D. Palle, Phys. Rev. D {\bf 43},
1585 (1991);
%
\bibitem{Gastao}P. Bicudo, G. Krein, J. Ribeiro and J. Villate,
Phys. Rev. D{\bf 45}, 1673 (1992);
%
\bibitem{meson}P. Bicudo and J. Ribeiro, Phys. Rev. D{\bf 42}, 1625 (1990);
%
\bibitem{prl} P. Bicudo Phys. Rev. Lett. {\bf 72}, 1600 (1994).
%
\bibitem{Lagae}
J. Lagae, Phys.Rev. D{\bf 45}, 317 (1992); A. Szczepaniak and
E. Swanson, Phys.Rev. D{\bf 55} 3987 (1997);
P. Bicudo, N. Brambilla, E. Ribeiro, A. Vairo, Phys. Lett.
B{\bf 442} 349 (1998). 
%
\bibitem{Hecht}
Fujiwara and Hecht, Nucl. Phys. A451, 625 (1986);
%
\bibitem{Goldman}F. Wang, G. Wu, L. Teng,
and T. Goldman, Phys. Rev. Lett. 69, 2891 (1992); Phys.
Rev. C {\bf 53}, 1161 (1996).
%
\bibitem{Isgur}P. Geiger and N. Isgur, Phys. Rev. D {\bf 44}, 799 (1991);
Phys. Rev. Lett. {\bf 67}, 1066 (1991);
Phys. Rev. D {\bf 47}, 5050 (1993).
%
\bibitem{Schrieffer} J. Schrieffer, '' Theory of Superconductivity '',
W. A. Bemjamin (1964);
%
\bibitem{Pagels} H. Pagels, Phys. Rev. D {\bf 14}, 2747 (1976).
%
\bibitem{Hooft}
G. 't Hooft Phys.\ Rev.\ Lett.\ {\bf 37}, 8 (1976);
Phys.\ Rev.\ D.\ {\bf 14}, 3432 (1976).
V. Bernard, R. L.Jaffe, U.-G. Meissner, Nucl. Phys. B {\bf 308},
753 (1988).
%
\end{references}
\end{document}